\documentclass[aps, amsmath, amssymb, reprint, superscriptaddress]{revtex4-1}

\usepackage{graphicx}
\usepackage{dcolumn}
\usepackage{bm}
\usepackage{amssymb}
\usepackage{subfigure}
\usepackage{soul}

\begin{document}

\preprint{Submitted to Physical Review Letters}

\title{Proof of concept for an ultrasensitive meta-device to detect and localize nonlinear elastic sources}

\author{M. Miniaci}
\affiliation{University of Le Havre, Laboratoire Ondes et Milieux Complexes, UMR CNRS 6294, 75 Rue Bellot, 76600 Le Havre, France}


\author{A. S. Gliozzi}
\email{antonio.gliozzi@polito.it}
\affiliation{Department of Applied Science and Technology, Politecnico di Torino, Corso Duca degli Abruzzi 24, 10129 Torino, Italy}

\author{B. Morvan}
\affiliation{University of Le Havre, Laboratoire Ondes et Milieux Complexes, UMR CNRS 6294, 75 Rue Bellot, 76600 Le Havre, France}

\author{A. Krushynska}%
\affiliation{Department of Physics, University of Torino, Via Pietro Giuria 1, 10125 Torino, Italy}

\author{F. Bosia}%
\affiliation{Department of Physics, University of Torino, Via Pietro Giuria 1, 10125 Torino, Italy}

\author{M. Scalerandi}%
\affiliation{Department of Applied Science and Technology, Politecnico di Torino, Corso Duca degli Abruzzi 24, 10129 Torino, Italy}

\author{N. M. Pugno}%
\altaffiliation[Also at: ]{School of Engineering and Materials Science, Queen Mary University of London, Mile End Road, London E1 4NS, United Kingdom}
\affiliation{Laboratory of Bio-Inspired and Graphene Nanomechanics, Department of Civil, Environmental and Mechanical Engineering, University of Trento, Via Mesiano 77, 38123 Trento, Italy}

\date{\today}

\begin{abstract}

The appearance of nonlinear effects in elastic wave propagation is one of the most reliable and sensitive indicators of the onset of material damage. However, these effects are usually very small and can be detected only using cumbersome digital signal processing techniques. Here, we propose and experimentally validate an alternative approach, using the filtering and focusing properties of elastic metamaterials to naturally select the higher harmonics generated by nonlinear effects and to increase their signal-to noise ratios, enabling the realization of time-reversal procedures for nonlinear elastic source detection. The proposed device demonstrates its potential as an efficient, compact, portable, passive apparatus for nonlinear elastic wave sensing and damage detection. 

\end{abstract}


\maketitle


In recent years, elastic metamaterials have attracted great attention due to their unconventional dynamic behaviour, with effects such as negative refraction \cite{Morvan_APL_2010}, frequency band gaps \cite{Kushwaha1993, Martinez_Sala_NATURE}, wave filtering/focusing \cite{Yang2004, Brun_JMPS_2010, Gliozzi_APL_2015}, acoustic cloaking \cite{Zhang2011, Sounas2015PRA, Kan2015PRA}, subwavelength sensing \cite{Zhang2016PRA, Zhu2016PRA}, etc. Their periodic structure, rather than single material constituents, is responsible for their behaviour, which exploits Bragg scattering \cite{deymier2013acoustic, Pennec2010229} or the presence of localized resonances \cite{craster2012acoustic, krushynska2014towards}. 
The attractive property of metamaterials to act as stop-band filters \cite{deymier2013acoustic} or to concentrate energy in selected frequency ranges \cite{carrara2013metamaterial} makes them potentially interesting for nonlinear elastic source detection and to reveal the presence of defects or cracks in a sample. This is because in general a nonlinear response is generated at the defect location and several possible features may appear, including the generation of higher order harmonics \cite{VanDenAbeele2000, Zaitsev_PRL_2014, Payan2007} or subharmonics \cite{Solodov2004, Ohara2007}, the nonlinear dependence of the elastic modulus and of attenuation coefficients on strain \cite{Renaud2009, Finkel2009, Trarieux2014} and, as a consequence, the shift of the resonance frequency with increasing excitation amplitude \cite{Scalerandi_APL_2016, Chen2011} and the failure of the superposition principle \cite{Scalerandi2008, Scalerandi2013}. All these possible signatures can be used to detect and monitor the presence and evolution of damage, exploiting the greater sensitivity of nonlinear detection techniques compared to conventional linear ones \cite{Ouarabi_PRB_2016}.

In the past years, nonlinear imaging techniques such as b-scan, c-scan or tomography \cite{Krohn2004} have attracted much interest. A particularly robust and efficient approach is Time Reversal (TR) and Nonlinear Elastic Wave Spectroscopy (TR-NEWS). This technique exploits space-time focusing of the wave-field achieved in TR \cite{ficek1997time} and applies it to a defect acting as a source of nonlinear elastic waves \cite{ulrich2006, gliozzi2006, ulrich2008time, Prada2002}. The scattered signal is recorded, the frequency generated by the primary source is filtered out using a band pass filter, and the resulting signal is time reversed and re-injected by the receiver: due to the ($t \rightarrow -t$) symmetry, the wavefield back-propagates to its original (nonlinear) source, focusing energy at the defect location at a specific time. Many studies have proved the efficiency of TR-NEWS in various configurations and for different types of nonlinear sources \cite{Goursolle2008170, ulrich2007, Zumpano20073666}, and numerical studies have quantified the robustness of the procedure in various experimental situations \cite{gliozzi2006}. However, most of the techniques for both detection and location of damage rely on extensive signal manipulation (normally, digital filtering) which might be critical in the case of short signals and/or when continuous signal acquisition is required (such as in acoustic emissions). Furthermore, the nonlinear components of the wavefield are often very small, if not submerged by the noise level, making it difficult to detect and estimate them. The concept adopted in this work overcomes these limitations combining TR-NEWS and phononic crystals in order to design a technique capable of naturally filtering out and/or concentrating energy in target frequency ranges. We experimentally demonstrate the feasibility and the efficiency of this technique, providing the proof of concept for ultra-sensitive metamaterial devices to detect and localize nonlinear elastic sources such as cracks or delaminations.

\begin{figure}
\centering
\begin{minipage}[]{0.68\linewidth}
\subfigure[]
{\includegraphics[trim=0mm 80mm 0mm 85mm, clip=true, width=1\textwidth]{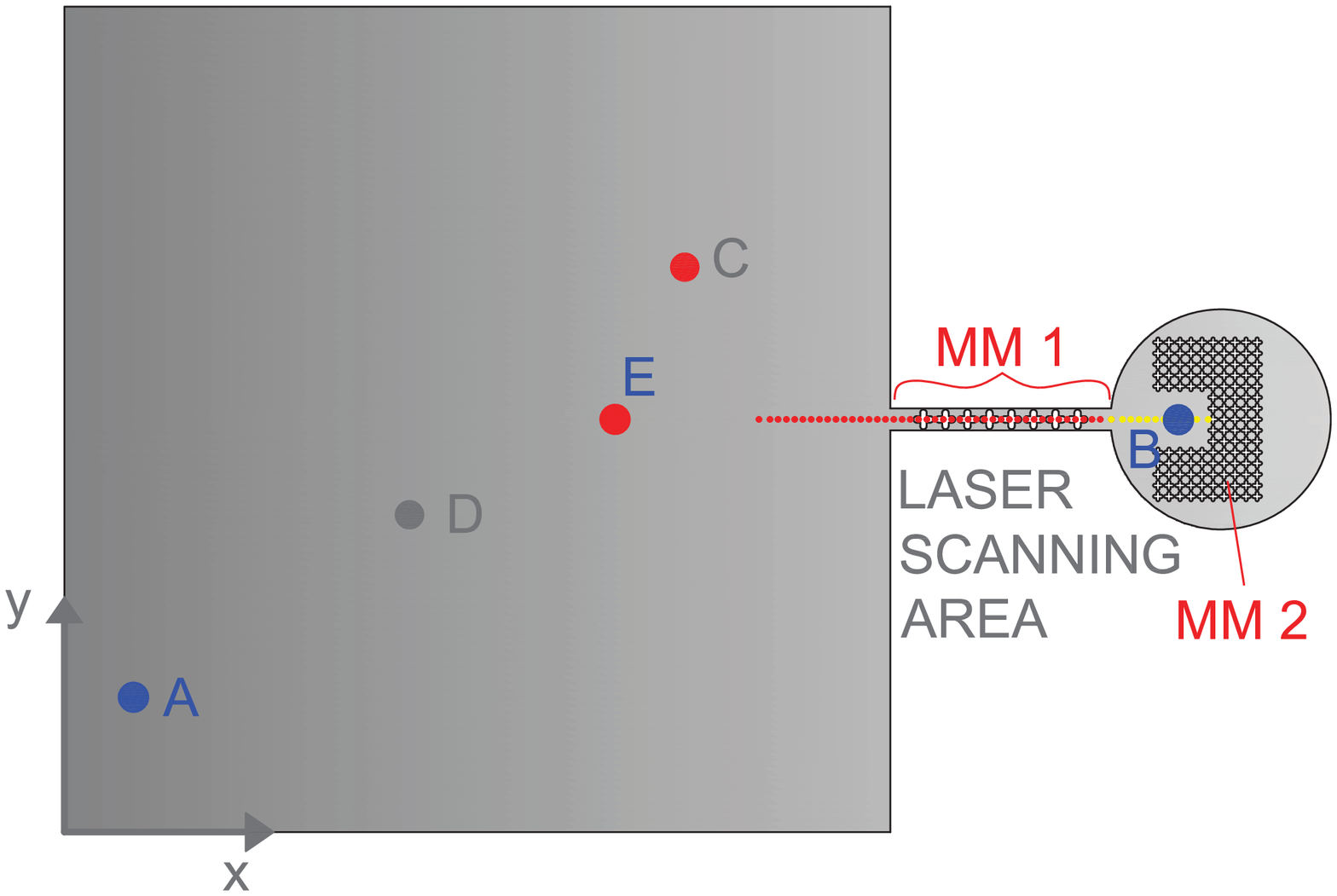}}
\end{minipage}
\begin{minipage}[]{0.30\linewidth}
\subfigure[]
{\includegraphics[trim=15mm 160mm 15mm 25mm, clip=true, width=1\textwidth]{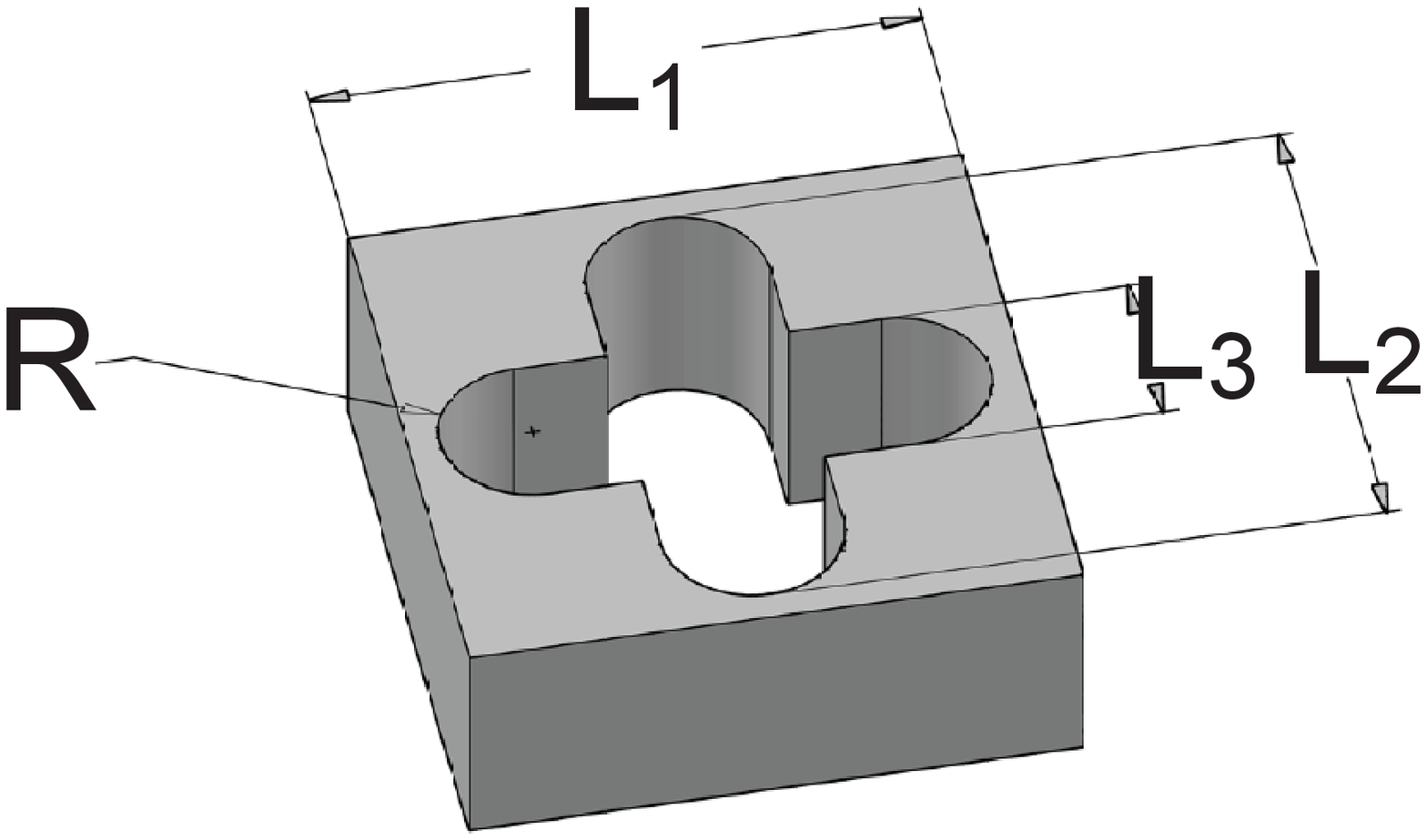}}
\subfigure[]
{\includegraphics[trim=15mm 180mm 67mm 25mm, clip=true, width=1\textwidth]{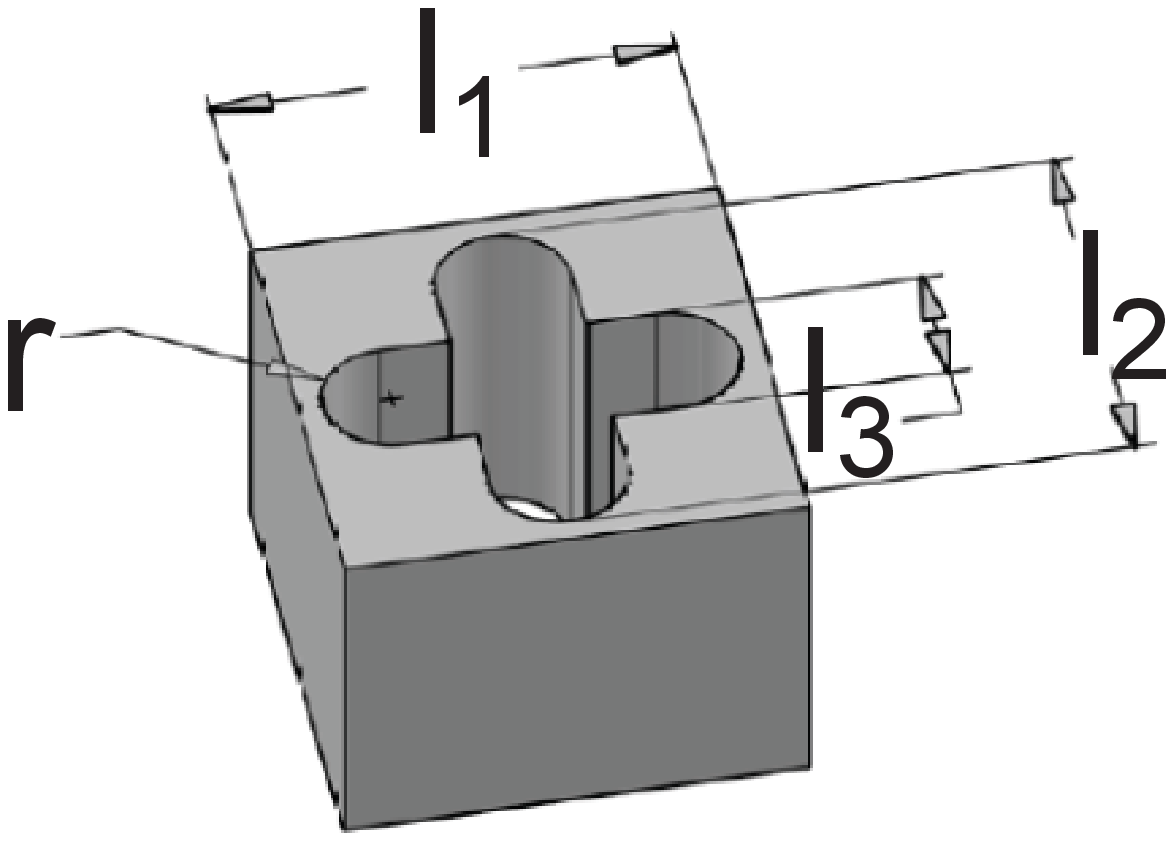}}
\end{minipage}
\caption{(a) Schematic representation of the specimen composed of an aluminium plate connected to a filtering metamaterial region (MM1) and a chaotic cavity \cite{BouMatar2009} with an additional focusing metamaterial-based structure (MM2). Three-dimensional view of the unit cells for (b) the MM1 region and (c) MM2 region.}
\label{fig1}
\end{figure}

A schematic representation of the experimental setup is given in Fig.\ref{fig1}. The sample is a pristine, undamaged $300 \times 300 \times 3$ mm$^3$ aluminium plate ($\rho = 2700$ kg$/$m$^3$, $E = 70$ GPa and $\nu = 0.33$) attached to which is a so-called "meta-device", consisting of two metamaterial regions (referred to as MM1 and MM2, respectively). MM1 consists of a 1D array of 8 cross-like cavities cut in a narrow rectangular waveguide and MM2 of a C-shaped array of smaller unit cells. The cavities are fabricated using waterjet cutting, with different lattice parameters, depending on the filtering (Fig.\ref{fig1}b) or reflecting (Fig.\ref{fig1}c) function they are designed for. Specifically, the lattice parameters are $L_1 = 8$ mm, $L_2 = 0.8 \cdot L_1$, $L_3 = 2R = 0.4 \cdot L_1$ for the MM1 region and $l_1 = 4$ mm, $l_2 = 0.8 \cdot l_1$, $l_3 = 2r = 0.4 \cdot l_1$ for the MM2 region, respectively. A cross-like geometry is chosen because it allows large Band Gap (BG) nucleation \cite{Miniaci_Ultrasonics_PVC}. 

\begin{figure}
\centering
\begin{minipage}[]{1\linewidth}
\subfigure[]
{\includegraphics[trim=0mm 0mm 76mm 0mm, clip=true, width=.47\linewidth]{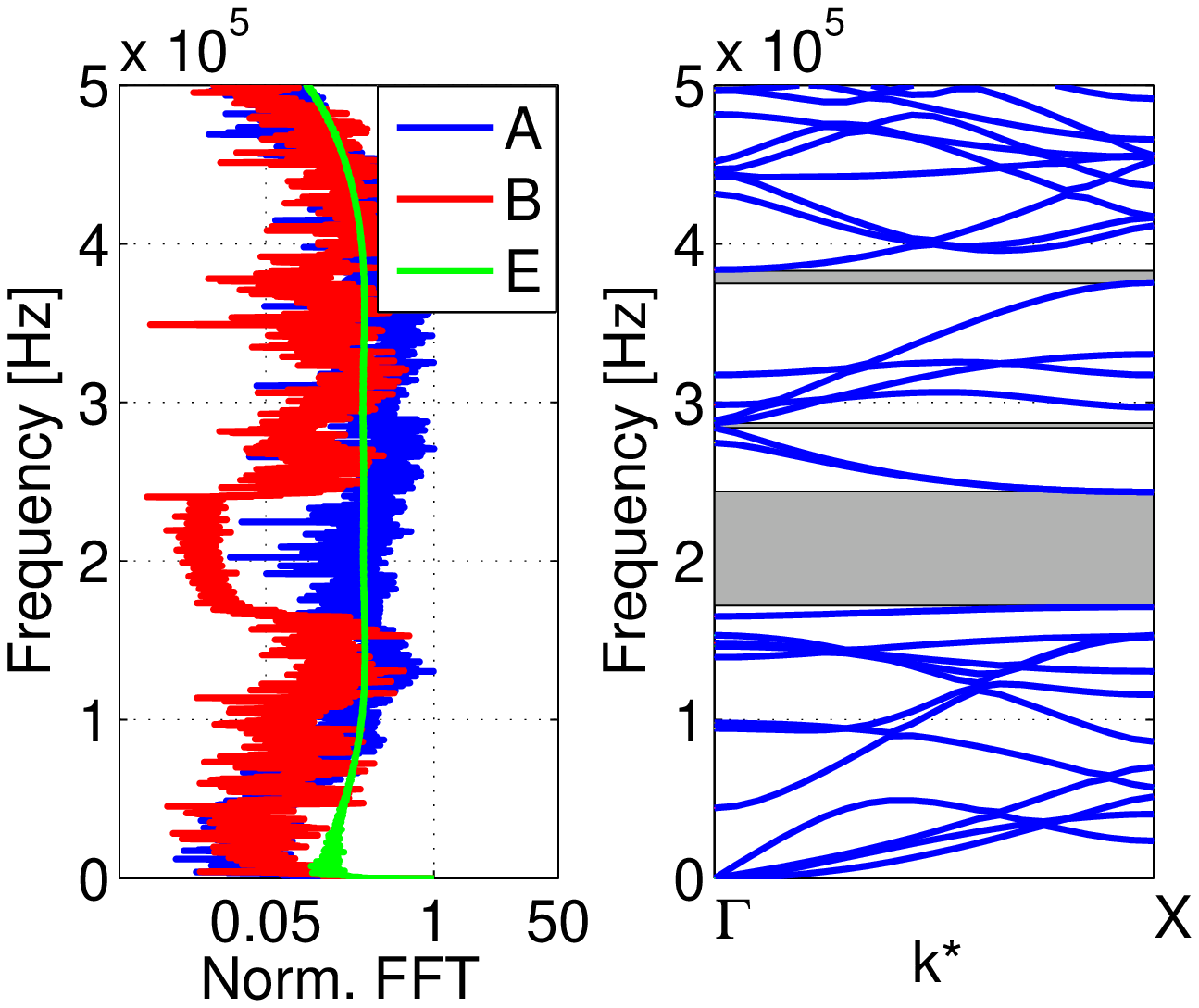}}
\subfigure[]
{\includegraphics[trim=72mm 0mm 0mm 0mm, clip=true, width=.495\linewidth]{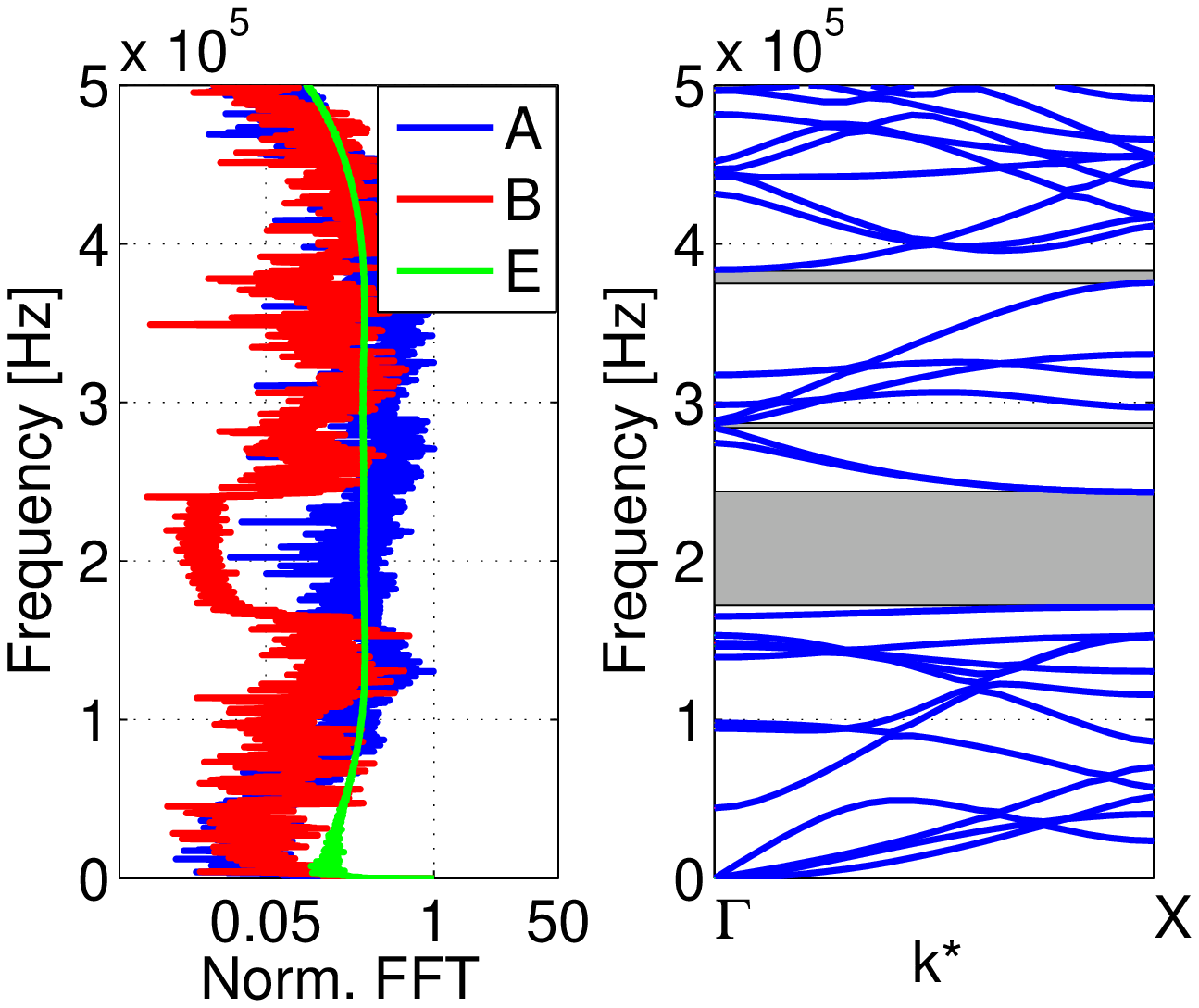}}
\end{minipage}
\caption{(a) Experimental normalized frequency spectrum for the MM1 region and (b) corresponding numerically-predicted dispersion band structure.}
\label{fig2}
\end{figure}

In order to investigate the BG structure of the MM1 region, its transmission spectrum is first investigated in a pitch-catch experiment \cite{Miniaci_Ultrasonics_PVC} according to the schematic representation of Fig. \ref{fig1}a (see \cite{REFSM} for details). An ultrasonic pulse with a frequency content between 50 and 450 kHz is launched by a transducer attached to the top surface of the plate (point E in Fig.\ref{fig1}a) and received at points A and B by means of 5 mm-diameter piezoelectric disk sensors. Fig.\ref{fig2}a shows the Fast Fourier Transform (FFT) of the input signal in E (green line) and those recorded in A and B (blue and red lines, respectively). A large BG appears between $172$ kHz and $244$ kHz, highlighted by a considerable frequency drop at the corresponding frequencies (up to $100$ dB). Two smaller BGs are visible around $285\,\mathrm{kHz}$ and $380\,\mathrm{kHz}$. On the contrary, the spectral content of the signal recorded in the plate (blue line in Fig.\ref{fig2}a), which is not subject to any filtering, shows the same frequency content as the excitation. These results are confirmed by numerically computed dispersion diagrams using Bloch-Floquet theory \cite{Collet_Ruzzene} in full 3D FEM simulations, shown in Fig.\ref{fig2}b. Here, the band structures are shown in terms of the reduced wave vector $k^* = [k_x \cdot L_1/ \pi; k_y \cdot L_1/ \pi]$ varying along the first irreducible Brillouin zone boundary $\Gamma-X$. Three BGs are predicted (highlighted in grey) in the considered frequency range, in excellent agreement with experimental results, thus validating the numerical approach.

\begin{figure}
\centering
\begin{minipage}[]{1\linewidth}
\subfigure[]
{\includegraphics[trim=0mm 0mm 0mm 0mm, clip=true, width=1\textwidth]{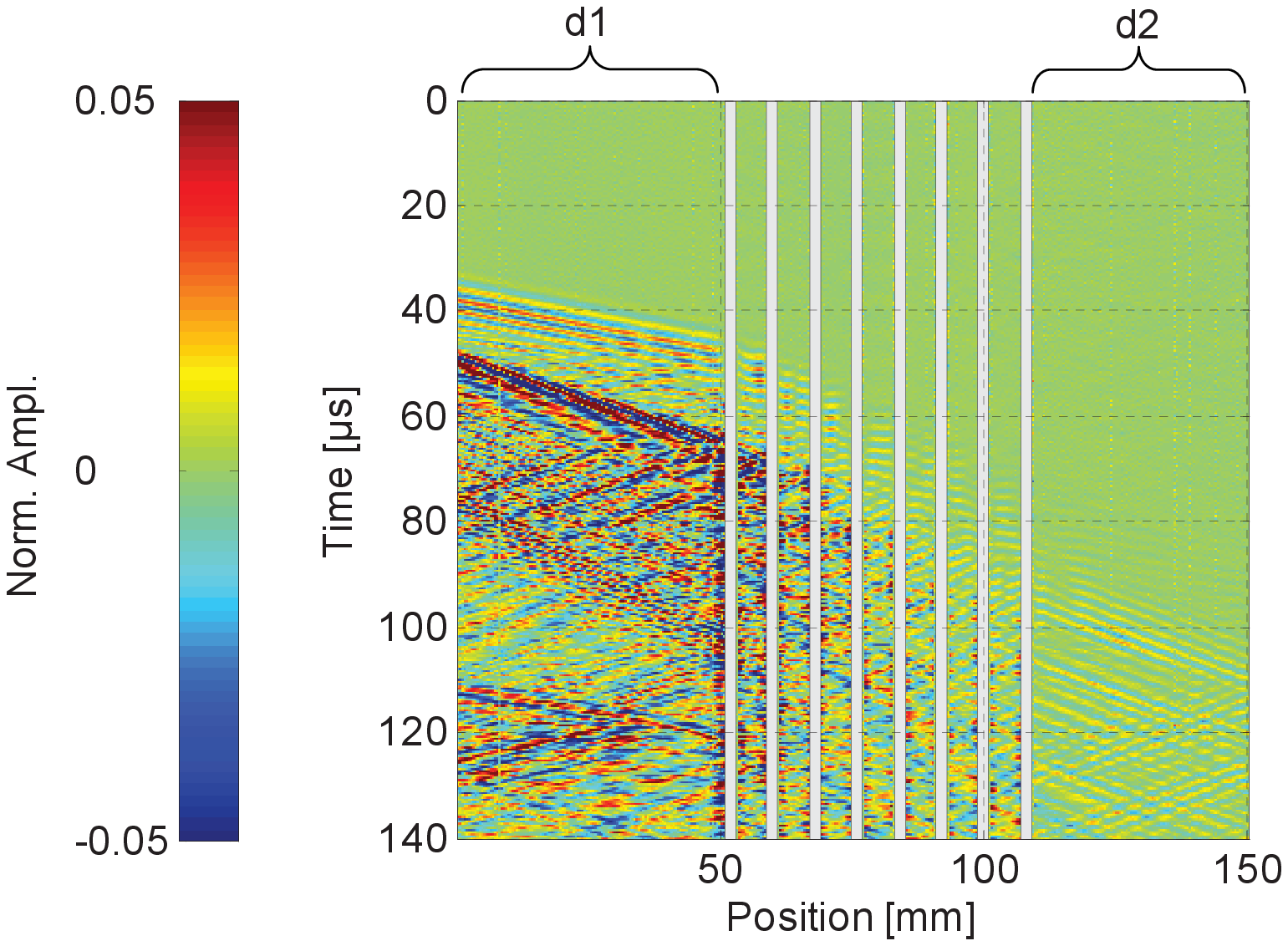}}
\end{minipage}
\begin{minipage}[]{1\linewidth}
\subfigure[]
{\includegraphics[trim=0mm 0mm 0mm 0mm, clip=true, width=1\textwidth]{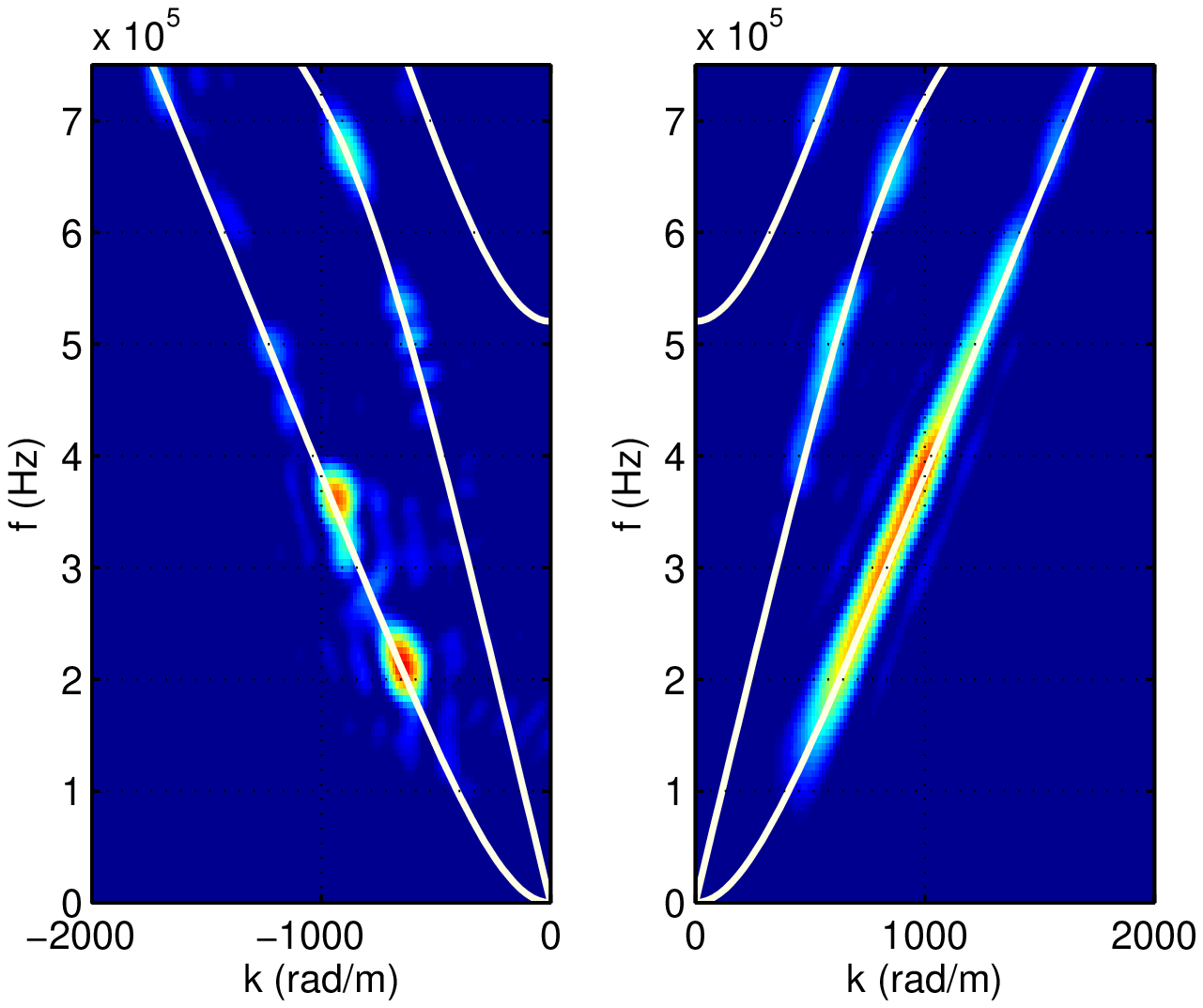}}
\end{minipage}
\caption{(a) Out-of-plane displacement as a function of time $t$ and position $d$ along the dotted line in Fig. \ref{fig1}a. White areas correspond to the cavities of the MM1 region. (t1, t2, t3, t4) are the time intervals used for signal processing of Fig. \ref{fig3}b. (b) Frequency $f$ vs. wavenumber $k$ representation of the measured signals. Theoretical dispersion curves (in white) are superimposed.}
\label{fig3}
\end{figure}

Further information of the dynamical properties of the MM1 region is provided by injecting a short pulse at point E and using a Scanning Laser Doppler Vibrometer (SLDV) to measure the out-of-plane component of the velocity at the surface of the plate along the dotted path highlighted in Fig. \ref{fig1}a (details are provided in \cite{REFSM}). Results for the space/time evolution of the measured amplitudes are shown in Figure \ref{fig3}a. Due to the antisymmetric excitation, mainly antisymmetric $A0$ Lamb waves are generated. However, Fig. \ref{fig3}a also clearly shows the presence of the reflected symmetric $S0$ mode (i.e. the faster waves visible in the $t=40 \mu s$ region). Strong reflections of the incident waves are clearly visible at a distance $d = 50$ mm from the source (corresponding to the first cavity in the MM1 waveguide) due to the impedance mismatch. Weaker reflected waves traveling through MM1 are also present at each cavity (white areas in the figure). Finally, the time vs. space representation shows a slight variation of the slope for the modes crossing the MM1 region, corresponding to a gradual decrease of the wave speed with the distance travelled in the MM1 for the different modes.

The signals detected at various positions along the path are processed by applying a two-dimensional Fourier transform (2D FFT) and determining the energy values for each processed point. This enables to obtain a frequency-wavenumber representation (Fig. \ref{fig3}b). Data is shown for the $d1$ acquisition region of the plate (i.e. from $d = 0$ up to the first cavity), with negative and positive values of the wavenumber $k_x$ providing information about reflected and incident waves, respectively. This representation clearly identifies the energy distribution among the excited modes. The energy maxima of the reflected waves (left panel) occur near the predicted BG frequency range, i.e. around $200$ kHz and $380$ kHz, associated with the incident antisymmetric fundamental mode $A0$ (right panel). Several supplementary areas with smaller amplitudes can be also observed in the vicinity of the $S0$ mode, deriving from direct generation by the transducer and/or from mode conversion occurring at the plate boundaries. Numerically predicted dispersion curves (white lines) are superimposed onto the experimental data, showing excellent agreement (further details and analysis relative to the $d2$ region are provided in \cite{REFSM}).

The role of the MM1 region is thus to act as a natural filter for the frequencies contained in the $172$ - $244$ kHz range. The excitation of a nonlinear material with a monochromatic wave of frequency contained in the BG of MM1 (e.g. $200$ kHz) produces higher harmonics in the plate, which are able to cross the MM1 barrier and to enter the circular "chaotic cavity" \cite{BouMatar2009}. Due to its ergodic properties and negligible absorption, the latter is widely used in TR experiments to generate multiple reflections and a reverberant acoustic field, making a single transducer sufficient for signal acquisition \cite{Draeger1997, SutinJASA2004, BouMatar2009}. The additional C-shaped metamaterial structure is designed to reflect the higher harmonics of the signal falling within a BG and to concentrate them in the geometric centre of the mirror, thus enhancing their signal-to-noise ratio. Analysis of SLDV-measured signals filtered at the frequencies of interest shows that there is good energy concentration at the centre of the mirror structure compared to peripheral regions in the chaotic cavity (\cite{REFSM}).

Additional FEM transmission simulations using the ABAQUS software are performed. The incoming wave is the superposition of two quasi-monocromatic waves centred around $f_1=200\,\mathrm{kHz}$ and $f_2=400\,\mathrm{kHz}$, respectively, with an imposed out-of-plane displacement of $1 \times 10^{-6}$ mm at point E. Two models are compared: one comprising both the filtering and the focusing regions and another consisting of only the filtering region (i.e. with a homogeneous aluminium chaotic cavity). 

\begin{figure}
\centering
\begin{minipage}[]{1\linewidth}
\subfigure[]
{\includegraphics[trim=0mm 0mm 0mm 0mm, clip=true, width=.48\textwidth]{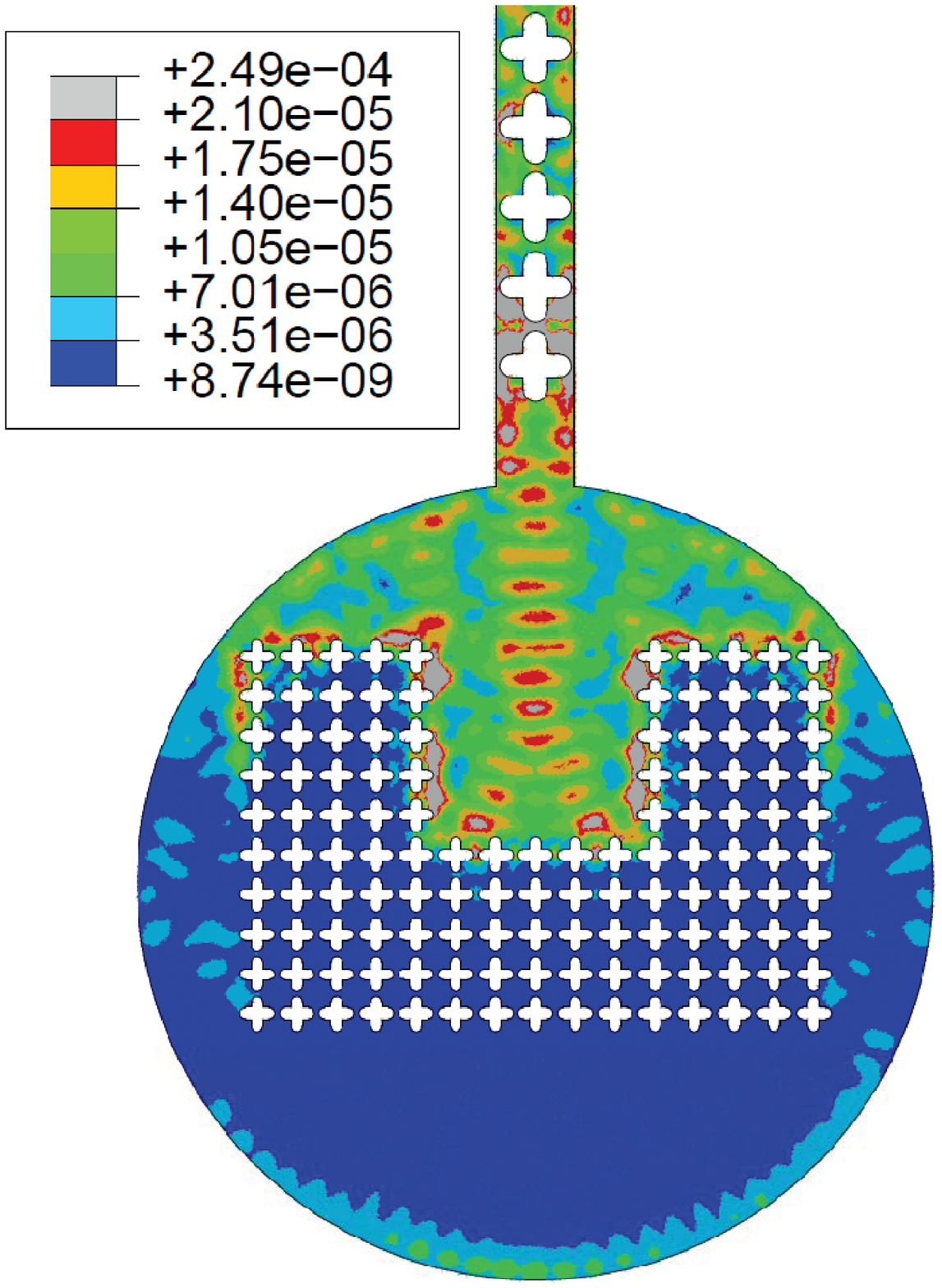}}
\subfigure[]
{\includegraphics[trim=0mm 0mm 0mm 0mm, clip=true, width=.48\textwidth]{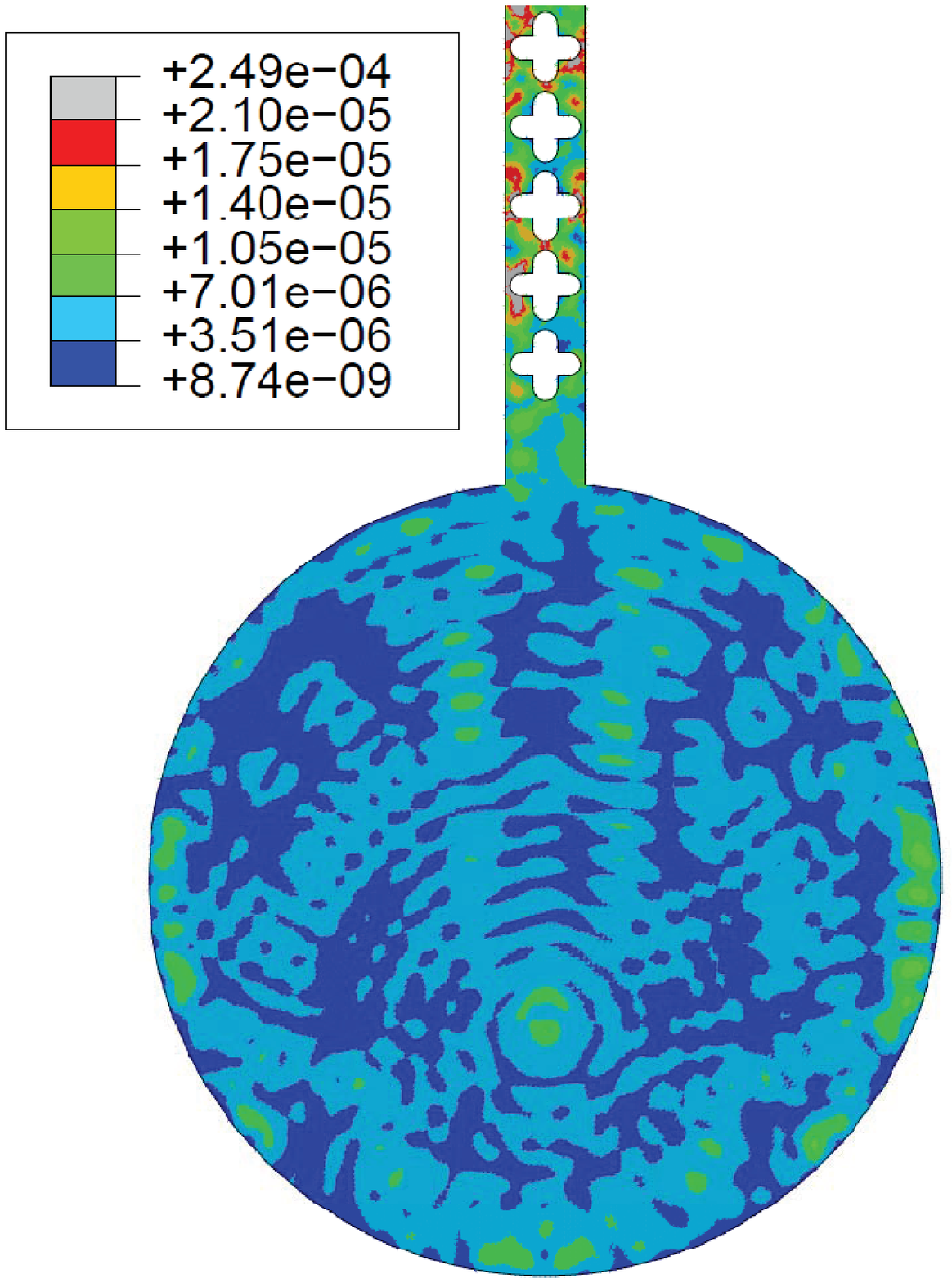}}
\end{minipage}
\caption{Snapshots of the Von-Mises stress (in MPa) inside the chaotic cavity showing the energy focusing by (a) the C-shaped phononic structure in comparison with (b) a homogeneous chaotic cavity at $t = 160$ $\mu$s.}
\label{fig4}
\end{figure}

Figs. \ref{fig4}a and \ref{fig4}b provide snapshots of the Von Mises stress maps at $t = 16 \cdot 10^{-5}$ s for the two configurations. In the case of a chaotic cavity with the C-shape metastructure, the formation of stationary waves occurs between the vertical portion of the mirror and the beginning of the waveguide. This allows the energy corresponding to the second harmonic (i.e. the signature of the nonlinearity) to be enhanced.

The possibility of combining both the filtering and focusing functionalities of the meta-device for NEWS-TR is now demonstrated. As discussed, the MM1 region provides a natural and selective filter that is transparent only to the higher harmonics generated by the nonlinear source and does not require any post-processing procedure. This allows the signal recorded by the sensor in the C-shaped mirror to be readily inverted and retransmitted into the sample. On the other hand, the MM2 region allows to focus energy of the second harmonic in order to enhance the signal-to-noise ratio in TR without the need for FFT filtering as in standard NEWS-TR applications, where the goal is to focus energy onto a nonlinear scatterer localized inside the sample.

\begin{figure}
\centering
\begin{minipage}[]{1\linewidth}
\subfigure[]
{\includegraphics[trim=0mm 0mm 0mm 0mm, clip=true, width=.49\textwidth]{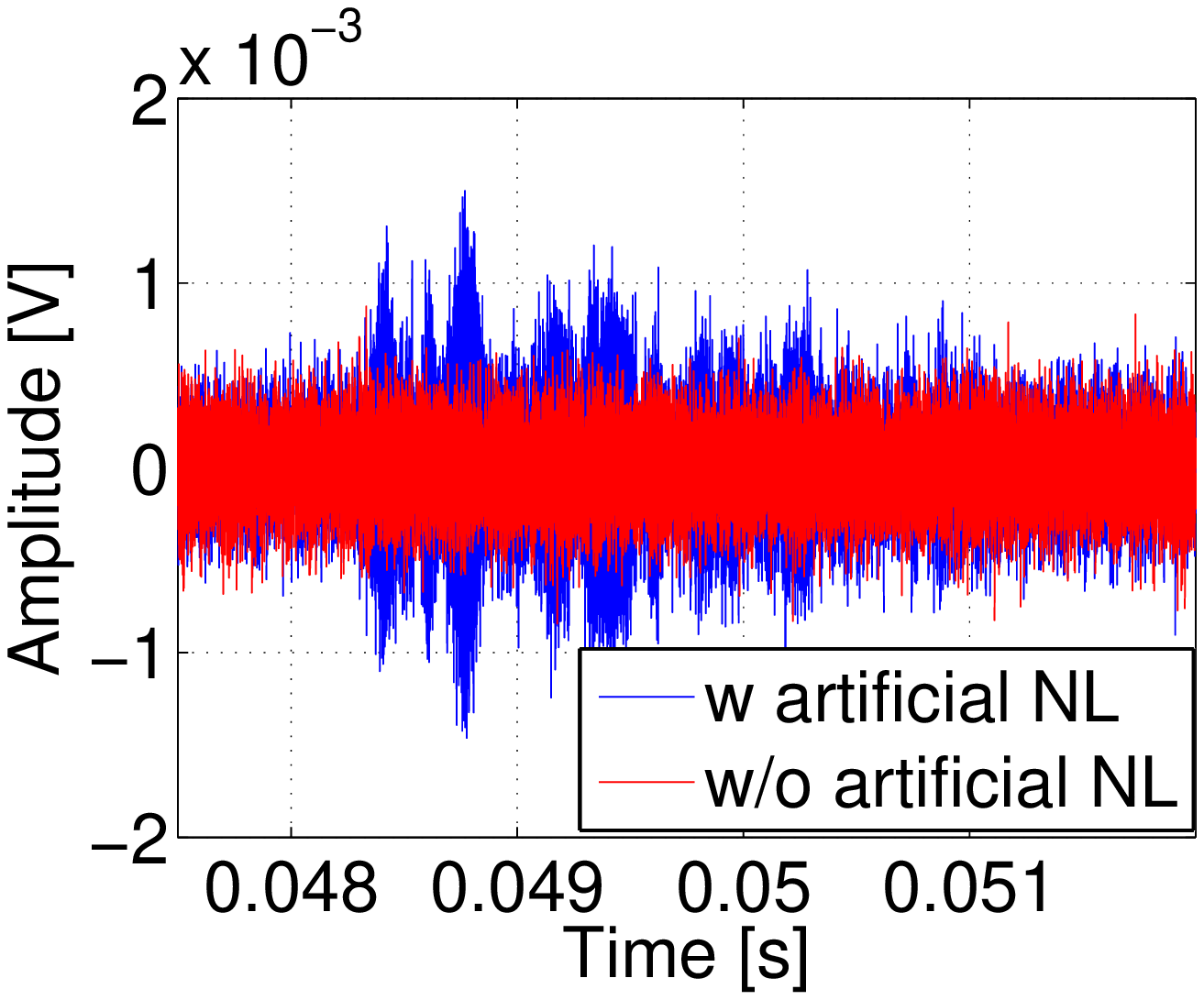}}
\subfigure[]
{\includegraphics[trim=0mm 0mm 0mm 0mm, clip=true, width=.49\textwidth]{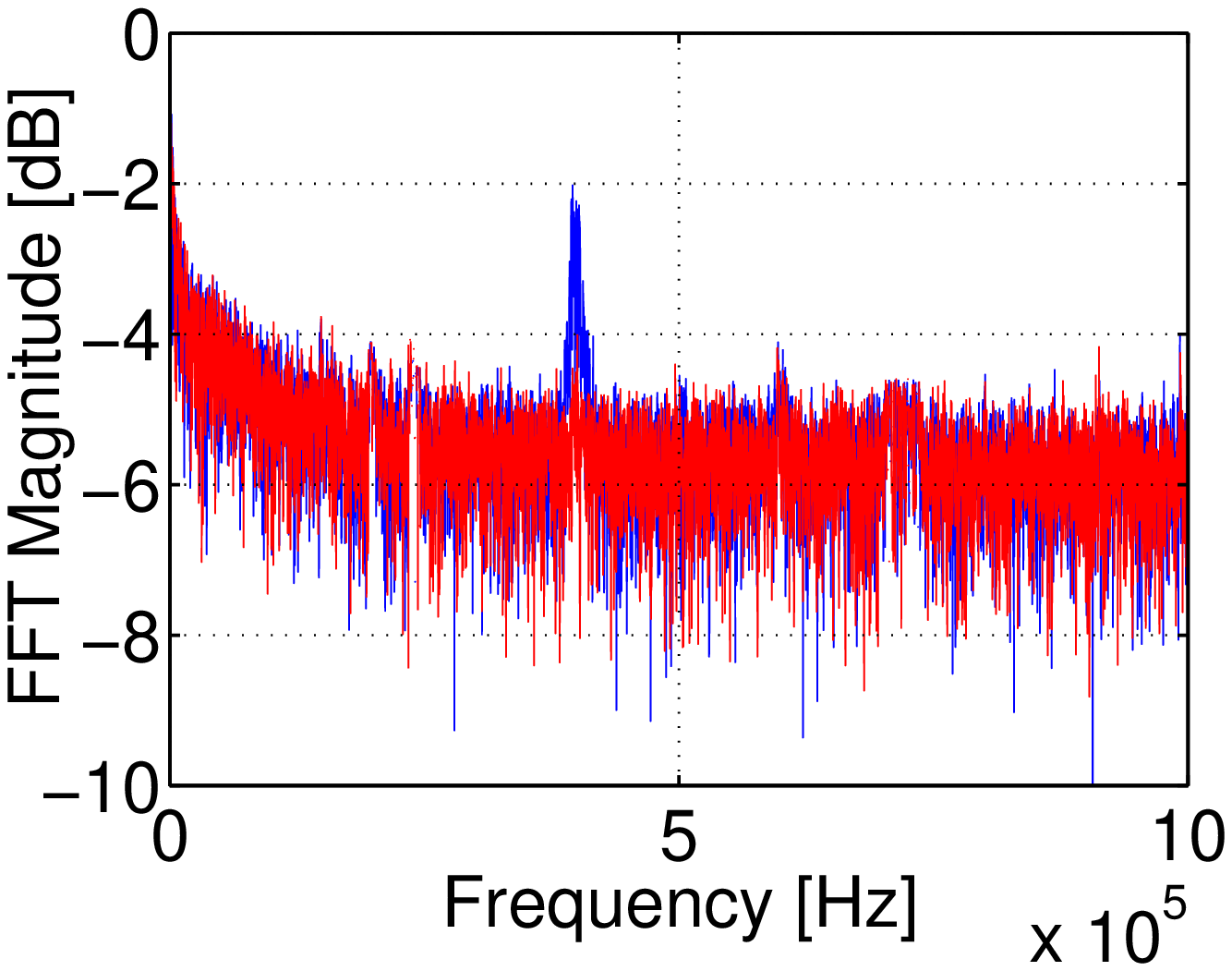}}
\subfigure[]
{\includegraphics[trim=0mm 0mm 0mm 0mm, clip=true, width=.49\textwidth]{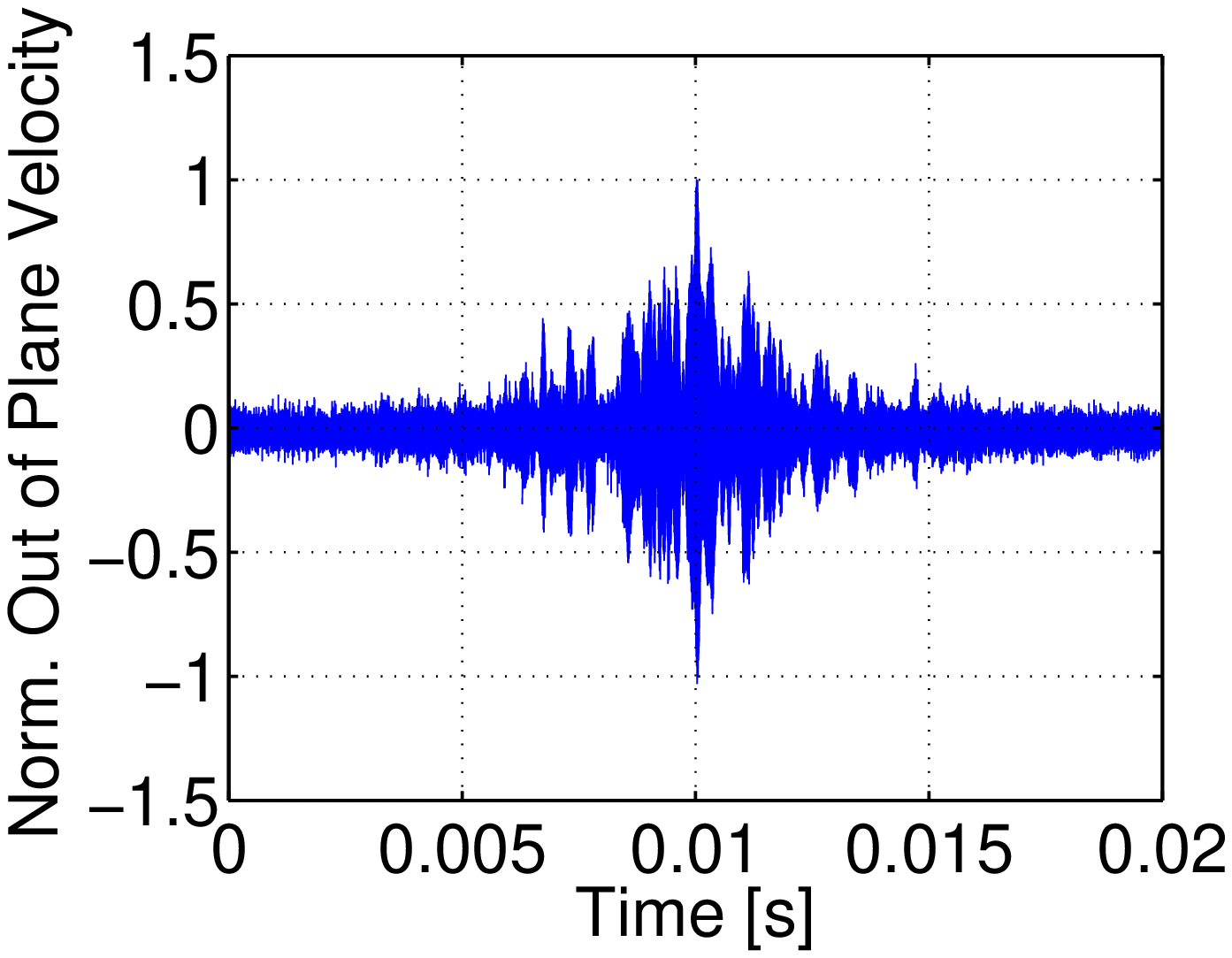}}
\subfigure[]
{\includegraphics[trim=0mm 0mm 0mm 0mm, clip=true, width=.49\textwidth]{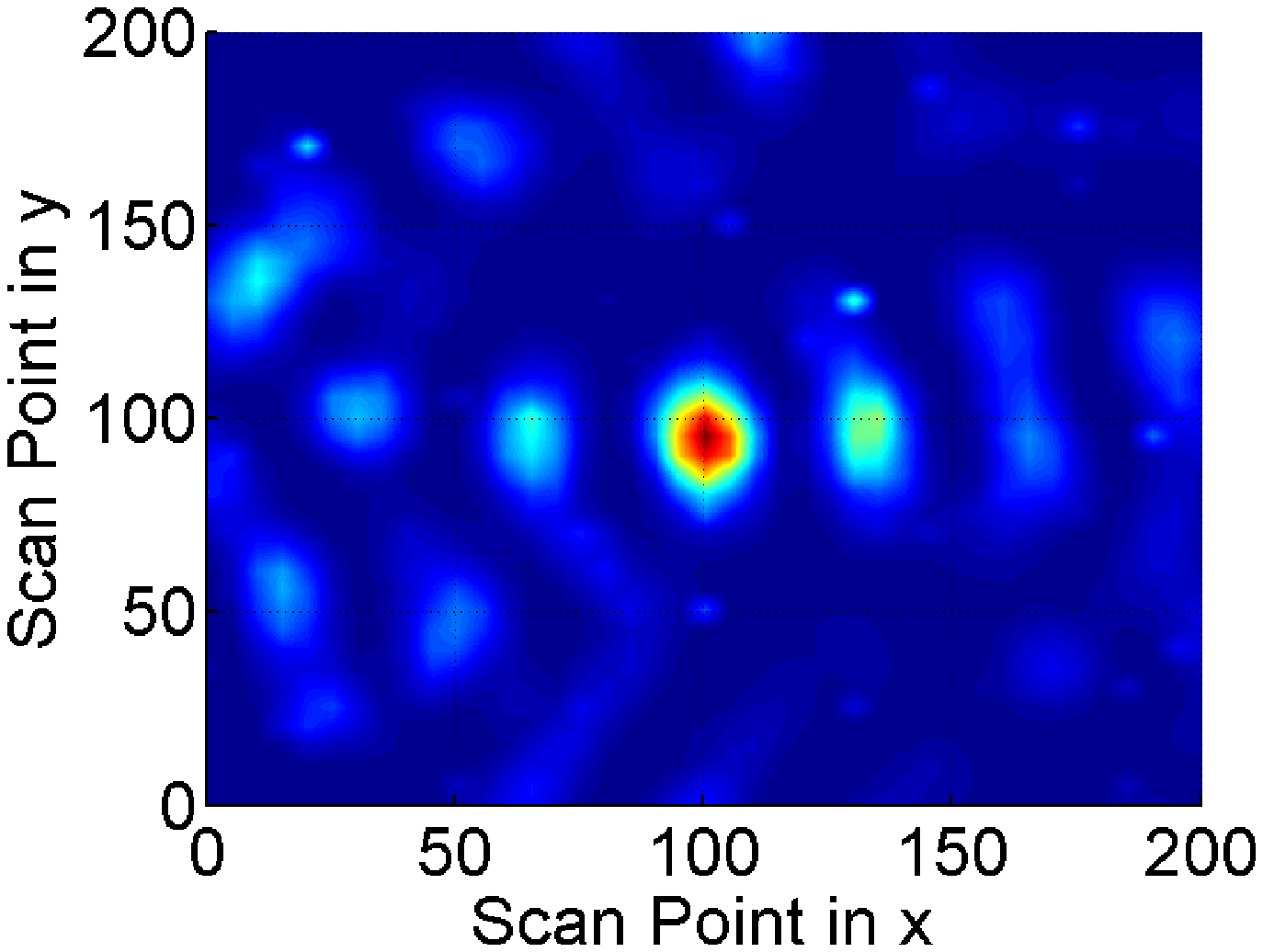}}
\end{minipage}
\caption{(a) Time and (b) frequency domain representations of the signals without (red) and with (blue) the nonlinear source acquired inside the C-shaped mirror. Note the difference in frequency content. Evidence of the (c) time and (d) spatial refocusing onto the nonlinear source location.}
\label{fig5}
\end{figure}

To perform the TR experiment, two piezoelectric disc transducers (PZT 1, $1.25$ cm$/1$ MHz, and PZT 2, $5$ mm) are placed on the plate at points C and B (Fig. \ref{fig1}a), the latter acting both as a receiver and as an actuator in the forward and backward TR propagation steps, respectively. Higher order harmonics are locally generated in a simple and reversible manner in the linear plate by inducing the wavefield to interact with an obstacle (e.g. a small cylinder, $8$ mm in diameter and $20$ mm in height, placed on the surface of the sample at a point D) whose contact surface has been previously humidified. The emergence of nonlinearity can be ascribed to two different reasons: the first is the fact that the two surfaces are in "clapping" contact \cite{Solodov2011},  mimicking the behaviour of a macroscopic crack in the sample; the second is due to the intrinsic nonlinear elastic behaviour of water \cite{Pecorari2006}, amplified by the presence of the small cylinder. The nonlinear behaviour is clearly visible in Figs. \ref{fig5}a  and \ref{fig5}b. The sample (with and without the nonlinear element) is excited by a source of the form
\begin{equation}
Y (t) = Y_1 = A_1\sin(2\pi f_1 t)*H(t_0)
\label{eq2}
\end{equation}

\noindent where $A_1$ is the amplitude of the sine function and $H(t_0)$ is the Hanning window centered in $t_0$ with a width corresponding to $21$ cycles of the sine wave of the fundamental frequency, $f_1$. The time and frequency domain representations of the signal received by PZT 2 are shown in Figs. \ref{fig5}a and \ref{fig5}b, respectively. In the case without the nonlinear scatterer, only noise is recorded inside the cavity (red line), whereas, in the presence of the artificial nonlinearity, a resonance peak appears (blue line) localized around the first harmonic of the fundamental frequency, (i.e. $f_2=2f_1= 400\,\mathrm{kHz}$). 


The NEWS-TR experiment is thus performed as follows: PZT 1 transducer emits a signal $Y=Y_1$ (Eq.(\ref{eq2})); the signal detected by PZT 2 (blue signal in Fig.\ref{fig5}a) is time reversed and transmitted back in the sample; SLDV measurements are performed on a spatial grid covering a $30\times 30\,\mathrm{mm}^2$ region around the nonlinear source (removed in the back propagation experiment) consisting of $200 \times 200$ equally spaced grid points. The laser vibrometer is positioned perpendicularly at $50$ cm from the surface to record the out-of-plane velocities of the points over the target area. Multiple ($128$) measurements are performed and averaged for each node, to filter out part of the noise.


After the backward propagation, time compression of the signal and spatial focusing of the wave field are observed at the nonlinear scatterer location. Fig. \ref{fig5}c reports an example of a time re-compressed signal detected with the SLDV. Also, at the focal time, the spatial map of the recorded velocities reveals focusing at the location of the defect, with a considerable concentration of energy, as shown in Fig. \ref{fig5}d. Notice that multiple scattering due to backward propagation of the wavefield through the metamaterial region increases the efficiency of TR and provides a high degree of focusing. Wavefield focalization on the nonlinear scatterer position occurs for amplitude ratios as small as $3\%$ between the generated nonlinear signal and the input \cite{REFSM}.

In conclusion, we have presented combined experimental and numerical results to demonstrate the feasibility of a novel passive sensor for signals generated by nonlinear elastic scatterers, such as cracks or delaminations. To do this, we have exploited the advanced frequency filtering and spatial focusing properties of elastic metamaterial structures, and proved the applicability of the sensor to time reversal experiments that allow to determine the spatial location of damage. The chosen metamaterial configurations, providing wide band gaps in the desired frequency ranges, were first characterized experimentally in pitch-catch experiments, demonstrating good agreement with numerical predictions based on FEM simulations. The dispersion properties of the metamaterial were also verified through scanning laser doppler vibrometer measurements. The same full-field non-contact technique was further adopted to confirm numerical predictions for wave filtering of the fundamental frequency by the first metamaterial region of the device. Higher harmonics were then generated by introducing a nonlinear source and focusing of the first harmonic in a chaotic cavity was demonstrated by means of a mirror-like metamaterial structure, confirming the possibility of extracting the signal above the noise level with no need for digital signal filtering. Finally, a time reversal experiment was carried out, showing good refocusing in time and space onto the nonlinear source, demonstrating the feasibility of the proposed meta-device for damage localization in structures.

In future, we aim to improve the design of this metamaterial-based sensor addressing issues such as optimized filter or focusing mirror designs, exploitation of multiple band gaps or frequency tunability using piezoelectric patches, and its effective application to external tested structures with reduced signal losses. Nevertheless, the results presented in this paper already provide the proof of concept for an efficient, portable, damage sensor with applications for passive continuous structural health and acoustic emission monitoring e.g. in civil engineering and the aerospace industry.

\vspace{1cm}
\textbf{Acknowledgments}

M.M. has received funding from the European Union's Horizon 2020 research and innovation programme under the Marie Sk{\l}odowska-Curie grant agreement N. 658483.
A.K. has received funding from the European Union's Seventh Framework programme for research and innovation under the Marie Sk{\l}odowska-Curie grant agreement N. 609402-2020 researchers: Train to Move (T2M).
N.M.P. is supported by the European Research Council (ERC StG Ideas 2011 BIHSNAM no. 279985 on "Bio-inspired hierarchical supernanomaterials" and ERC PoC 2015 SILKENE No. 693670), and by the European Commission under the Graphene FET Flagship (WP14 "Polymer Nanocomposites" No. 604391). FB is supported by BIHSNAM.

\vspace{1cm}


\begin{thebibliography}{46}%
\makeatletter
\providecommand \@ifxundefined [1]{%
 \@ifx{#1\undefined}
}%
\providecommand \@ifnum [1]{%
 \ifnum #1\expandafter \@firstoftwo
 \else \expandafter \@secondoftwo
 \fi
}%
\providecommand \@ifx [1]{%
 \ifx #1\expandafter \@firstoftwo
 \else \expandafter \@secondoftwo
 \fi
}%
\providecommand \natexlab [1]{#1}%
\providecommand \enquote  [1]{``#1''}%
\providecommand \bibnamefont  [1]{#1}%
\providecommand \bibfnamefont [1]{#1}%
\providecommand \citenamefont [1]{#1}%
\providecommand \href@noop [0]{\@secondoftwo}%
\providecommand \href [0]{\begingroup \@sanitize@url \@href}%
\providecommand \@href[1]{\@@startlink{#1}\@@href}%
\providecommand \@@href[1]{\endgroup#1\@@endlink}%
\providecommand \@sanitize@url [0]{\catcode `\\12\catcode `\$12\catcode
  `\&12\catcode `\#12\catcode `\^12\catcode `\_12\catcode `\%12\relax}%
\providecommand \@@startlink[1]{}%
\providecommand \@@endlink[0]{}%
\providecommand \url  [0]{\begingroup\@sanitize@url \@url }%
\providecommand \@url [1]{\endgroup\@href {#1}{\urlprefix }}%
\providecommand \urlprefix  [0]{URL }%
\providecommand \Eprint [0]{\href }%
\providecommand \doibase [0]{http://dx.doi.org/}%
\providecommand \selectlanguage [0]{\@gobble}%
\providecommand \bibinfo  [0]{\@secondoftwo}%
\providecommand \bibfield  [0]{\@secondoftwo}%
\providecommand \translation [1]{[#1]}%
\providecommand \BibitemOpen [0]{}%
\providecommand \bibitemStop [0]{}%
\providecommand \bibitemNoStop [0]{.\EOS\space}%
\providecommand \EOS [0]{\spacefactor3000\relax}%
\providecommand \BibitemShut  [1]{\csname bibitem#1\endcsname}%
\let\auto@bib@innerbib\@empty
\bibitem [{\citenamefont {Morvan}\ \emph {et~al.}(2010)\citenamefont {Morvan},
  \citenamefont {Tinel}, \citenamefont {Hladky-Hennion}, \citenamefont
  {Vasseur},\ and\ \citenamefont {Dubus}}]{Morvan_APL_2010}%
  \BibitemOpen
  \bibfield  {author} {\bibinfo {author} {\bibfnamefont {B.}~\bibnamefont
  {Morvan}}, \bibinfo {author} {\bibfnamefont {A.}~\bibnamefont {Tinel}},
  \bibinfo {author} {\bibfnamefont {A.-C.}\ \bibnamefont {Hladky-Hennion}},
  \bibinfo {author} {\bibfnamefont {J.}~\bibnamefont {Vasseur}}, \ and\
  \bibinfo {author} {\bibfnamefont {B.}~\bibnamefont {Dubus}},\ }\href@noop {}
  {\bibfield  {journal} {\bibinfo  {journal} {Applied Physics Letters}\
  }\textbf {\bibinfo {volume} {96}},\ \bibinfo {eid} {101905} (\bibinfo {year}
  {2010})}\BibitemShut {NoStop}%
\bibitem [{\citenamefont {Kushwaha}\ \emph {et~al.}(1993)\citenamefont
  {Kushwaha}, \citenamefont {Halevi}, \citenamefont {Dobrzynski},\ and\
  \citenamefont {Djafari-Rouhani}}]{Kushwaha1993}%
  \BibitemOpen
  \bibfield  {author} {\bibinfo {author} {\bibfnamefont {M.~S.}\ \bibnamefont
  {Kushwaha}}, \bibinfo {author} {\bibfnamefont {P.}~\bibnamefont {Halevi}},
  \bibinfo {author} {\bibfnamefont {L.}~\bibnamefont {Dobrzynski}}, \ and\
  \bibinfo {author} {\bibfnamefont {B.}~\bibnamefont {Djafari-Rouhani}},\
  }\href@noop {} {\bibfield  {journal} {\bibinfo  {journal} {Physical Review
  Letters}\ }\textbf {\bibinfo {volume} {71}},\ \bibinfo {pages} {2022}
  (\bibinfo {year} {1993})}\BibitemShut {NoStop}%
\bibitem [{\citenamefont {Martinez-Sala}\ \emph {et~al.}(1995)\citenamefont
  {Martinez-Sala}, \citenamefont {Sancho}, \citenamefont {Sanchez},
  \citenamefont {Gomez}, \citenamefont {Llinares},\ and\ \citenamefont
  {Meseguer}}]{Martinez_Sala_NATURE}%
  \BibitemOpen
  \bibfield  {author} {\bibinfo {author} {\bibfnamefont {R.}~\bibnamefont
  {Martinez-Sala}}, \bibinfo {author} {\bibfnamefont {J.}~\bibnamefont
  {Sancho}}, \bibinfo {author} {\bibfnamefont {J.~V.}\ \bibnamefont {Sanchez}},
  \bibinfo {author} {\bibfnamefont {V.}~\bibnamefont {Gomez}}, \bibinfo
  {author} {\bibfnamefont {J.}~\bibnamefont {Llinares}}, \ and\ \bibinfo
  {author} {\bibfnamefont {F.}~\bibnamefont {Meseguer}},\ }\href@noop {}
  {\bibfield  {journal} {\bibinfo  {journal} {Nature}\ }\textbf {\bibinfo
  {volume} {378}},\ \bibinfo {pages} {241 } (\bibinfo {year}
  {1995})}\BibitemShut {NoStop}%
\bibitem [{\citenamefont {Yang}\ \emph {et~al.}(2004)\citenamefont {Yang},
  \citenamefont {Page}, \citenamefont {Liu}, \citenamefont {Cowan},
  \citenamefont {Chan},\ and\ \citenamefont {Sheng}}]{Yang2004}%
  \BibitemOpen
  \bibfield  {author} {\bibinfo {author} {\bibfnamefont {S.}~\bibnamefont
  {Yang}}, \bibinfo {author} {\bibfnamefont {J.~H.}\ \bibnamefont {Page}},
  \bibinfo {author} {\bibfnamefont {Z.}~\bibnamefont {Liu}}, \bibinfo {author}
  {\bibfnamefont {M.~L.}\ \bibnamefont {Cowan}}, \bibinfo {author}
  {\bibfnamefont {C.~T.}\ \bibnamefont {Chan}}, \ and\ \bibinfo {author}
  {\bibfnamefont {P.}~\bibnamefont {Sheng}},\ }\href@noop {} {\bibfield
  {journal} {\bibinfo  {journal} {Physical Review Letters}\ }\textbf {\bibinfo
  {volume} {93}},\ \bibinfo {pages} {024301} (\bibinfo {year}
  {2004})}\BibitemShut {NoStop}%
\bibitem [{\citenamefont {Brun}\ \emph {et~al.}(2010)\citenamefont {Brun},
  \citenamefont {Guenneau}, \citenamefont {Movchan},\ and\ \citenamefont
  {Bigoni}}]{Brun_JMPS_2010}%
  \BibitemOpen
  \bibfield  {author} {\bibinfo {author} {\bibfnamefont {M.}~\bibnamefont
  {Brun}}, \bibinfo {author} {\bibfnamefont {S.}~\bibnamefont {Guenneau}},
  \bibinfo {author} {\bibfnamefont {A.~B.}\ \bibnamefont {Movchan}}, \ and\
  \bibinfo {author} {\bibfnamefont {D.}~\bibnamefont {Bigoni}},\ }\href@noop {}
  {\bibfield  {journal} {\bibinfo  {journal} {Journal of the Mechanics and
  Physics of Solids}\ }\textbf {\bibinfo {volume} {58}},\ \bibinfo {pages}
  {1212 } (\bibinfo {year} {2010})}\BibitemShut {NoStop}%
\bibitem [{\citenamefont {Gliozzi}\ \emph {et~al.}(2015)\citenamefont
  {Gliozzi}, \citenamefont {Miniaci}, \citenamefont {Bosia}, \citenamefont
  {Pugno},\ and\ \citenamefont {Scalerandi}}]{Gliozzi_APL_2015}%
  \BibitemOpen
  \bibfield  {author} {\bibinfo {author} {\bibfnamefont {A.~S.}\ \bibnamefont
  {Gliozzi}}, \bibinfo {author} {\bibfnamefont {M.}~\bibnamefont {Miniaci}},
  \bibinfo {author} {\bibfnamefont {F.}~\bibnamefont {Bosia}}, \bibinfo
  {author} {\bibfnamefont {N.~M.}\ \bibnamefont {Pugno}}, \ and\ \bibinfo
  {author} {\bibfnamefont {M.}~\bibnamefont {Scalerandi}},\ }\href@noop {}
  {\bibfield  {journal} {\bibinfo  {journal} {Applied Physics Letters}\
  }\textbf {\bibinfo {volume} {107}},\ \bibinfo {eid} {161902} (\bibinfo {year}
  {2015})}\BibitemShut {NoStop}%
\bibitem [{\citenamefont {Zhang}\ \emph {et~al.}(2011)\citenamefont {Zhang},
  \citenamefont {Xia},\ and\ \citenamefont {Fang}}]{Zhang2011}%
  \BibitemOpen
  \bibfield  {author} {\bibinfo {author} {\bibfnamefont {S.}~\bibnamefont
  {Zhang}}, \bibinfo {author} {\bibfnamefont {C.}~\bibnamefont {Xia}}, \ and\
  \bibinfo {author} {\bibfnamefont {N.}~\bibnamefont {Fang}},\ }\href@noop {}
  {\bibfield  {journal} {\bibinfo  {journal} {Physical Review Letters}\
  }\textbf {\bibinfo {volume} {106}},\ \bibinfo {pages} {024301} (\bibinfo
  {year} {2011})}\BibitemShut {NoStop}%
\bibitem [{\citenamefont {Sounas}\ \emph {et~al.}(2015)\citenamefont {Sounas},
  \citenamefont {Fleury},\ and\ \citenamefont {Al\`u}}]{Sounas2015PRA}%
  \BibitemOpen
  \bibfield  {author} {\bibinfo {author} {\bibfnamefont {D.~L.}\ \bibnamefont
  {Sounas}}, \bibinfo {author} {\bibfnamefont {R.}~\bibnamefont {Fleury}}, \
  and\ \bibinfo {author} {\bibfnamefont {A.}~\bibnamefont {Al\`u}},\
  }\href@noop {} {\bibfield  {journal} {\bibinfo  {journal} {Physical Review
  Applied}\ }\textbf {\bibinfo {volume} {4}},\ \bibinfo {eid} {014005}
  (\bibinfo {year} {2015})}\BibitemShut {NoStop}%
\bibitem [{\citenamefont {Kan}\ \emph {et~al.}(2015)\citenamefont {Kan},
  \citenamefont {Garc\'ia-Chocano}, \citenamefont {Cervera}, \citenamefont
  {Liang}, \citenamefont {Zou}, \citenamefont {Yin}, \citenamefont {Cheng},\
  and\ \citenamefont {S\'anchez-Dehesa}}]{Kan2015PRA}%
  \BibitemOpen
  \bibfield  {author} {\bibinfo {author} {\bibfnamefont {W.}~\bibnamefont
  {Kan}}, \bibinfo {author} {\bibfnamefont {V.~M.}\ \bibnamefont
  {Garc\'ia-Chocano}}, \bibinfo {author} {\bibfnamefont {F.}~\bibnamefont
  {Cervera}}, \bibinfo {author} {\bibfnamefont {B.}~\bibnamefont {Liang}},
  \bibinfo {author} {\bibfnamefont {X.-y.}\ \bibnamefont {Zou}}, \bibinfo
  {author} {\bibfnamefont {L.-l.}\ \bibnamefont {Yin}}, \bibinfo {author}
  {\bibfnamefont {J.}~\bibnamefont {Cheng}}, \ and\ \bibinfo {author}
  {\bibfnamefont {J.}~\bibnamefont {S\'anchez-Dehesa}},\ }\href@noop {}
  {\bibfield  {journal} {\bibinfo  {journal} {Physical Review Applied}\
  }\textbf {\bibinfo {volume} {3}},\ \bibinfo {eid} {064019} (\bibinfo {year}
  {2015})}\BibitemShut {NoStop}%
\bibitem [{\citenamefont {Zhang}\ \emph {et~al.}(2016)\citenamefont {Zhang},
  \citenamefont {Zhang}, \citenamefont {Guo}, \citenamefont {Leng},
  \citenamefont {Feng},\ and\ \citenamefont {Cao}}]{Zhang2016PRA}%
  \BibitemOpen
  \bibfield  {author} {\bibinfo {author} {\bibfnamefont {S.}~\bibnamefont
  {Zhang}}, \bibinfo {author} {\bibfnamefont {Y.}~\bibnamefont {Zhang}},
  \bibinfo {author} {\bibfnamefont {Y.}~\bibnamefont {Guo}}, \bibinfo {author}
  {\bibfnamefont {Y.}~\bibnamefont {Leng}}, \bibinfo {author} {\bibfnamefont
  {W.}~\bibnamefont {Feng}}, \ and\ \bibinfo {author} {\bibfnamefont
  {W.}~\bibnamefont {Cao}},\ }\href@noop {} {\bibfield  {journal} {\bibinfo
  {journal} {Physical Review Applied}\ }\textbf {\bibinfo {volume} {5}},\
  \bibinfo {eid} {034006} (\bibinfo {year} {2016})}\BibitemShut {NoStop}%
\bibitem [{\citenamefont {Zhu}\ \emph {et~al.}(2016)\citenamefont {Zhu},
  \citenamefont {Liang}, \citenamefont {Kan}, \citenamefont {Peng},\ and\
  \citenamefont {Cheng}}]{Zhu2016PRA}%
  \BibitemOpen
  \bibfield  {author} {\bibinfo {author} {\bibfnamefont {X.}~\bibnamefont
  {Zhu}}, \bibinfo {author} {\bibfnamefont {B.}~\bibnamefont {Liang}}, \bibinfo
  {author} {\bibfnamefont {W.}~\bibnamefont {Kan}}, \bibinfo {author}
  {\bibfnamefont {Y.}~\bibnamefont {Peng}}, \ and\ \bibinfo {author}
  {\bibfnamefont {J.}~\bibnamefont {Cheng}},\ }\href@noop {} {\bibfield
  {journal} {\bibinfo  {journal} {Physical Review Applied}\ }\textbf {\bibinfo
  {volume} {5}},\ \bibinfo {eid} {054015} (\bibinfo {year} {2016})}\BibitemShut
  {NoStop}%
\bibitem [{\citenamefont {Deymier}(2013)}]{deymier2013acoustic}%
  \BibitemOpen
  \bibfield  {author} {\bibinfo {author} {\bibfnamefont {P.}~\bibnamefont
  {Deymier}},\ }\href@noop {} {\emph {\bibinfo {title} {Acoustic Metamaterials
  and Phononic Crystals}}},\ Springer Series in Solid-State Sciences\ (\bibinfo
   {publisher} {Springer Berlin Heidelberg},\ \bibinfo {year}
  {2013})\BibitemShut {NoStop}%
\bibitem [{\citenamefont {Pennec}\ \emph {et~al.}(2010)\citenamefont {Pennec},
  \citenamefont {Vasseur}, \citenamefont {Djafari-Rouhani}, \citenamefont
  {Dobrzy\'nski},\ and\ \citenamefont {Deymier}}]{Pennec2010229}%
  \BibitemOpen
  \bibfield  {author} {\bibinfo {author} {\bibfnamefont {Y.}~\bibnamefont
  {Pennec}}, \bibinfo {author} {\bibfnamefont {J.~O.}\ \bibnamefont {Vasseur}},
  \bibinfo {author} {\bibfnamefont {B.}~\bibnamefont {Djafari-Rouhani}},
  \bibinfo {author} {\bibfnamefont {L.}~\bibnamefont {Dobrzy\'nski}}, \ and\
  \bibinfo {author} {\bibfnamefont {P.~A.}\ \bibnamefont {Deymier}},\
  }\href@noop {} {\bibfield  {journal} {\bibinfo  {journal} {Surface Science
  Reports}\ }\textbf {\bibinfo {volume} {65}},\ \bibinfo {pages} {229 }
  (\bibinfo {year} {2010})}\BibitemShut {NoStop}%
\bibitem [{\citenamefont {Craster}\ and\ \citenamefont
  {Guenneau}(2012)}]{craster2012acoustic}%
  \BibitemOpen
  \bibfield  {author} {\bibinfo {author} {\bibfnamefont {R.}~\bibnamefont
  {Craster}}\ and\ \bibinfo {author} {\bibfnamefont {S.}~\bibnamefont
  {Guenneau}},\ }\href@noop {} {\emph {\bibinfo {title} {Acoustic
  Metamaterials: Negative Refraction, Imaging, Lensing and Cloaking}}},\
  Springer Series in Materials Science\ (\bibinfo  {publisher} {Springer
  London, Limited},\ \bibinfo {year} {2012})\BibitemShut {NoStop}%
\bibitem [{\citenamefont {Krushynska}\ \emph {et~al.}(2014)\citenamefont
  {Krushynska}, \citenamefont {Kouznetsova},\ and\ \citenamefont
  {Geers}}]{krushynska2014towards}%
  \BibitemOpen
  \bibfield  {author} {\bibinfo {author} {\bibfnamefont {A.}~\bibnamefont
  {Krushynska}}, \bibinfo {author} {\bibfnamefont {V.}~\bibnamefont
  {Kouznetsova}}, \ and\ \bibinfo {author} {\bibfnamefont {M.}~\bibnamefont
  {Geers}},\ }\href@noop {} {\bibfield  {journal} {\bibinfo  {journal} {Journal
  of the Mechanics and Physics of Solids}\ }\textbf {\bibinfo {volume} {71}},\
  \bibinfo {pages} {179} (\bibinfo {year} {2014})}\BibitemShut {NoStop}%
\bibitem [{\citenamefont {Carrara}\ \emph {et~al.}(2013)\citenamefont
  {Carrara}, \citenamefont {Cacan}, \citenamefont {Toussaint}, \citenamefont
  {Leamy}, \citenamefont {Ruzzene},\ and\ \citenamefont
  {Erturk}}]{carrara2013metamaterial}%
  \BibitemOpen
  \bibfield  {author} {\bibinfo {author} {\bibfnamefont {M.}~\bibnamefont
  {Carrara}}, \bibinfo {author} {\bibfnamefont {M.}~\bibnamefont {Cacan}},
  \bibinfo {author} {\bibfnamefont {J.}~\bibnamefont {Toussaint}}, \bibinfo
  {author} {\bibfnamefont {M.}~\bibnamefont {Leamy}}, \bibinfo {author}
  {\bibfnamefont {M.}~\bibnamefont {Ruzzene}}, \ and\ \bibinfo {author}
  {\bibfnamefont {A.}~\bibnamefont {Erturk}},\ }\href@noop {} {\bibfield
  {journal} {\bibinfo  {journal} {Smart Materials and Structures}\ }\textbf
  {\bibinfo {volume} {22}},\ \bibinfo {pages} {065004} (\bibinfo {year}
  {2013})}\BibitemShut {NoStop}%
\bibitem [{\citenamefont {Van Den~Abeele}\ \emph {et~al.}(2000)\citenamefont
  {Van Den~Abeele}, \citenamefont {Johnson},\ and\ \citenamefont
  {Sutin}}]{VanDenAbeele2000}%
  \BibitemOpen
  \bibfield  {author} {\bibinfo {author} {\bibfnamefont {K.-A.}\ \bibnamefont
  {Van Den~Abeele}}, \bibinfo {author} {\bibfnamefont {P.~A.}\ \bibnamefont
  {Johnson}}, \ and\ \bibinfo {author} {\bibfnamefont {A.}~\bibnamefont
  {Sutin}},\ }\href@noop {} {\bibfield  {journal} {\bibinfo  {journal}
  {Research in nondestructive evaluation}\ }\textbf {\bibinfo {volume} {12}},\
  \bibinfo {pages} {17} (\bibinfo {year} {2000})}\BibitemShut {NoStop}%
\bibitem [{\citenamefont {Zaitsev}\ \emph {et~al.}(2014)\citenamefont
  {Zaitsev}, \citenamefont {Gusev}, \citenamefont {Tournat},\ and\
  \citenamefont {Richard}}]{Zaitsev_PRL_2014}%
  \BibitemOpen
  \bibfield  {author} {\bibinfo {author} {\bibfnamefont {V.~Y.}\ \bibnamefont
  {Zaitsev}}, \bibinfo {author} {\bibfnamefont {V.~E.}\ \bibnamefont {Gusev}},
  \bibinfo {author} {\bibfnamefont {V.}~\bibnamefont {Tournat}}, \ and\
  \bibinfo {author} {\bibfnamefont {P.}~\bibnamefont {Richard}},\ }\href@noop
  {} {\bibfield  {journal} {\bibinfo  {journal} {Physical Review Letters}\
  }\textbf {\bibinfo {volume} {112}},\ \bibinfo {pages} {108302} (\bibinfo
  {year} {2014})}\BibitemShut {NoStop}%
\bibitem [{\citenamefont {Payan}\ \emph {et~al.}(2007)\citenamefont {Payan},
  \citenamefont {Garnier},\ and\ \citenamefont {Moysan}}]{Payan2007}%
  \BibitemOpen
  \bibfield  {author} {\bibinfo {author} {\bibfnamefont {C.}~\bibnamefont
  {Payan}}, \bibinfo {author} {\bibfnamefont {V.}~\bibnamefont {Garnier}}, \
  and\ \bibinfo {author} {\bibfnamefont {J.}~\bibnamefont {Moysan}},\
  }\href@noop {} {\bibfield  {journal} {\bibinfo  {journal} {Journal of
  Acoustical Sociecty of America}\ }\textbf {\bibinfo {volume} {121}},\
  \bibinfo {pages} {EL125} (\bibinfo {year} {2007})}\BibitemShut {NoStop}%
\bibitem [{\citenamefont {Solodov}\ \emph {et~al.}(2004)\citenamefont
  {Solodov}, \citenamefont {Wackerl},\ and\ \citenamefont
  {et~al.}}]{Solodov2004}%
  \BibitemOpen
  \bibfield  {author} {\bibinfo {author} {\bibfnamefont {I.}~\bibnamefont
  {Solodov}}, \bibinfo {author} {\bibfnamefont {J.}~\bibnamefont {Wackerl}}, \
  and\ \bibinfo {author} {\bibfnamefont {K.~P.}\ \bibnamefont {et~al.}},\
  }\href@noop {} {\bibfield  {journal} {\bibinfo  {journal} {Applied Physics
  Letters}\ }\textbf {\bibinfo {volume} {84}},\ \bibinfo {pages} {5386}
  (\bibinfo {year} {2004})}\BibitemShut {NoStop}%
\bibitem [{\citenamefont {Ohara}\ \emph {et~al.}(2007)\citenamefont {Ohara},
  \citenamefont {Mihara}, \citenamefont {Sasaki}, \citenamefont {Ogata},
  \citenamefont {Yamamoto}, \citenamefont {Kishimoto},\ and\ \citenamefont
  {Yamanaka}}]{Ohara2007}%
  \BibitemOpen
  \bibfield  {author} {\bibinfo {author} {\bibfnamefont {Y.}~\bibnamefont
  {Ohara}}, \bibinfo {author} {\bibfnamefont {T.}~\bibnamefont {Mihara}},
  \bibinfo {author} {\bibfnamefont {R.}~\bibnamefont {Sasaki}}, \bibinfo
  {author} {\bibfnamefont {T.}~\bibnamefont {Ogata}}, \bibinfo {author}
  {\bibfnamefont {S.}~\bibnamefont {Yamamoto}}, \bibinfo {author}
  {\bibfnamefont {Y.}~\bibnamefont {Kishimoto}}, \ and\ \bibinfo {author}
  {\bibfnamefont {K.}~\bibnamefont {Yamanaka}},\ }\href@noop {} {\bibfield
  {journal} {\bibinfo  {journal} {Applied Physics Letters}\ }\textbf {\bibinfo
  {volume} {90}},\ \bibinfo {eid} {011902} (\bibinfo {year}
  {2007})}\BibitemShut {NoStop}%
\bibitem [{\citenamefont {Renaud}\ \emph {et~al.}(2009)\citenamefont {Renaud},
  \citenamefont {CallÃ©},\ and\ \citenamefont {Defontaine}}]{Renaud2009}%
  \BibitemOpen
  \bibfield  {author} {\bibinfo {author} {\bibfnamefont {G.}~\bibnamefont
  {Renaud}}, \bibinfo {author} {\bibfnamefont {S.}~\bibnamefont {CallÃ©}}, \
  and\ \bibinfo {author} {\bibfnamefont {M.}~\bibnamefont {Defontaine}},\
  }\href@noop {} {\bibfield  {journal} {\bibinfo  {journal} {Applied Physics
  Letters}\ }\textbf {\bibinfo {volume} {94}},\ \bibinfo {pages} {011905}
  (\bibinfo {year} {2009})}\BibitemShut {NoStop}%
\bibitem [{\citenamefont {Finkel}\ \emph {et~al.}(2009)\citenamefont {Finkel},
  \citenamefont {Zhou}, \citenamefont {Basu}, \citenamefont {Yeheskel},\ and\
  \citenamefont {Barsoum}}]{Finkel2009}%
  \BibitemOpen
  \bibfield  {author} {\bibinfo {author} {\bibfnamefont {P.}~\bibnamefont
  {Finkel}}, \bibinfo {author} {\bibfnamefont {A.~G.}\ \bibnamefont {Zhou}},
  \bibinfo {author} {\bibfnamefont {S.}~\bibnamefont {Basu}}, \bibinfo {author}
  {\bibfnamefont {O.}~\bibnamefont {Yeheskel}}, \ and\ \bibinfo {author}
  {\bibfnamefont {M.~W.}\ \bibnamefont {Barsoum}},\ }\href@noop {} {\bibfield
  {journal} {\bibinfo  {journal} {Applied Physics Letters}\ }\textbf {\bibinfo
  {volume} {94}},\ \bibinfo {eid} {241904} (\bibinfo {year}
  {2009})}\BibitemShut {NoStop}%
\bibitem [{\citenamefont {Trarieux}\ \emph {et~al.}(2014)\citenamefont
  {Trarieux}, \citenamefont {Call\'e}, \citenamefont {Moreschi}, \citenamefont
  {Renaud},\ and\ \citenamefont {Defontaine}}]{Trarieux2014}%
  \BibitemOpen
  \bibfield  {author} {\bibinfo {author} {\bibfnamefont {C.}~\bibnamefont
  {Trarieux}}, \bibinfo {author} {\bibfnamefont {S.}~\bibnamefont {Call\'e}},
  \bibinfo {author} {\bibfnamefont {H.}~\bibnamefont {Moreschi}}, \bibinfo
  {author} {\bibfnamefont {G.}~\bibnamefont {Renaud}}, \ and\ \bibinfo {author}
  {\bibfnamefont {M.}~\bibnamefont {Defontaine}},\ }\href@noop {} {\bibfield
  {journal} {\bibinfo  {journal} {Applied Physics Letters}\ }\textbf {\bibinfo
  {volume} {105}},\ \bibinfo {pages} {264103} (\bibinfo {year}
  {2014})}\BibitemShut {NoStop}%
\bibitem [{\citenamefont {Scalerandi}\ \emph {et~al.}(2016)\citenamefont
  {Scalerandi}, \citenamefont {Gliozzi}, \citenamefont {Ouarabi},\ and\
  \citenamefont {Boubenider}}]{Scalerandi_APL_2016}%
  \BibitemOpen
  \bibfield  {author} {\bibinfo {author} {\bibfnamefont {M.}~\bibnamefont
  {Scalerandi}}, \bibinfo {author} {\bibfnamefont {A.~S.}\ \bibnamefont
  {Gliozzi}}, \bibinfo {author} {\bibfnamefont {M.~A.}\ \bibnamefont
  {Ouarabi}}, \ and\ \bibinfo {author} {\bibfnamefont {F.}~\bibnamefont
  {Boubenider}},\ }\href@noop {} {\bibfield  {journal} {\bibinfo  {journal}
  {Applied Physics Letters}\ }\textbf {\bibinfo {volume} {108}},\ \bibinfo
  {pages} {214103} (\bibinfo {year} {2016})}\BibitemShut {NoStop}%
\bibitem [{\citenamefont {Chen}\ \emph {et~al.}(2011)\citenamefont {Chen},
  \citenamefont {Kim}, \citenamefont {Kurtis},\ and\ \citenamefont
  {Jacobs}}]{Chen2011}%
  \BibitemOpen
  \bibfield  {author} {\bibinfo {author} {\bibfnamefont {J.}~\bibnamefont
  {Chen}}, \bibinfo {author} {\bibfnamefont {J.}~\bibnamefont {Kim}}, \bibinfo
  {author} {\bibfnamefont {K.}~\bibnamefont {Kurtis}}, \ and\ \bibinfo {author}
  {\bibfnamefont {L.}~\bibnamefont {Jacobs}},\ }\href@noop {} {\bibfield
  {journal} {\bibinfo  {journal} {Journal of Acoustical Sociecty of America}\
  }\textbf {\bibinfo {volume} {130}},\ \bibinfo {pages} {2728} (\bibinfo {year}
  {2011})}\BibitemShut {NoStop}%
\bibitem [{\citenamefont {Scalerandi}\ \emph {et~al.}(2008)\citenamefont
  {Scalerandi}, \citenamefont {Gliozzi}, \citenamefont {Bruno}, \citenamefont
  {Masera},\ and\ \citenamefont {Bocca}}]{Scalerandi2008}%
  \BibitemOpen
  \bibfield  {author} {\bibinfo {author} {\bibfnamefont {M.}~\bibnamefont
  {Scalerandi}}, \bibinfo {author} {\bibfnamefont {A.}~\bibnamefont {Gliozzi}},
  \bibinfo {author} {\bibfnamefont {C.~L.~E.}\ \bibnamefont {Bruno}}, \bibinfo
  {author} {\bibfnamefont {D.}~\bibnamefont {Masera}}, \ and\ \bibinfo {author}
  {\bibfnamefont {P.}~\bibnamefont {Bocca}},\ }\href@noop {} {\bibfield
  {journal} {\bibinfo  {journal} {Applied Physics Letters}\ }\textbf {\bibinfo
  {volume} {92}},\ \bibinfo {pages} {101912} (\bibinfo {year}
  {2008})}\BibitemShut {NoStop}%
\bibitem [{\citenamefont {Scalerandi}\ \emph {et~al.}(2013)\citenamefont
  {Scalerandi}, \citenamefont {Griffa}, \citenamefont {Antonaci}, \citenamefont
  {Wyrzykowski},\ and\ \citenamefont {Lura}}]{Scalerandi2013}%
  \BibitemOpen
  \bibfield  {author} {\bibinfo {author} {\bibfnamefont {M.}~\bibnamefont
  {Scalerandi}}, \bibinfo {author} {\bibfnamefont {M.}~\bibnamefont {Griffa}},
  \bibinfo {author} {\bibfnamefont {P.}~\bibnamefont {Antonaci}}, \bibinfo
  {author} {\bibfnamefont {M.}~\bibnamefont {Wyrzykowski}}, \ and\ \bibinfo
  {author} {\bibfnamefont {P.}~\bibnamefont {Lura}},\ }\href@noop {} {\bibfield
   {journal} {\bibinfo  {journal} {Journal of Applied Physics}\ }\textbf
  {\bibinfo {volume} {113}},\ \bibinfo {pages} {154902} (\bibinfo {year}
  {2013})}\BibitemShut {NoStop}%
\bibitem [{\citenamefont {Ouarabi}\ \emph {et~al.}(2016)\citenamefont
  {Ouarabi}, \citenamefont {Boubenider}, \citenamefont {Gliozzi},\ and\
  \citenamefont {Scalerandi}}]{Ouarabi_PRB_2016}%
  \BibitemOpen
  \bibfield  {author} {\bibinfo {author} {\bibfnamefont {M.~A.}\ \bibnamefont
  {Ouarabi}}, \bibinfo {author} {\bibfnamefont {F.}~\bibnamefont {Boubenider}},
  \bibinfo {author} {\bibfnamefont {A.~S.}\ \bibnamefont {Gliozzi}}, \ and\
  \bibinfo {author} {\bibfnamefont {M.}~\bibnamefont {Scalerandi}},\
  }\href@noop {} {\bibfield  {journal} {\bibinfo  {journal} {Physical Review
  B}\ }\textbf {\bibinfo {volume} {94}},\ \bibinfo {pages} {134103} (\bibinfo
  {year} {2016})}\BibitemShut {NoStop}%
\bibitem [{\citenamefont {Krohn}\ \emph {et~al.}(2004)\citenamefont {Krohn},
  \citenamefont {Pfleiderer}, \citenamefont {Stoessel}, \citenamefont
  {Solodov},\ and\ \citenamefont {Busse}}]{Krohn2004}%
  \BibitemOpen
  \bibfield  {author} {\bibinfo {author} {\bibfnamefont {N.}~\bibnamefont
  {Krohn}}, \bibinfo {author} {\bibfnamefont {K.}~\bibnamefont {Pfleiderer}},
  \bibinfo {author} {\bibfnamefont {R.}~\bibnamefont {Stoessel}}, \bibinfo
  {author} {\bibfnamefont {I.}~\bibnamefont {Solodov}}, \ and\ \bibinfo
  {author} {\bibfnamefont {G.}~\bibnamefont {Busse}},\ }\href@noop {}
  {\bibfield  {journal} {\bibinfo  {journal} {Acoustical Imaging}\ }\textbf
  {\bibinfo {volume} {27}},\ \bibinfo {pages} {91} (\bibinfo {year}
  {2004})}\BibitemShut {NoStop}%
\bibitem [{\citenamefont {Ficek}\ and\ \citenamefont
  {Drummond}(1997)}]{ficek1997time}%
  \BibitemOpen
  \bibfield  {author} {\bibinfo {author} {\bibfnamefont {Z.}~\bibnamefont
  {Ficek}}\ and\ \bibinfo {author} {\bibfnamefont {P.~D.}\ \bibnamefont
  {Drummond}},\ }\href@noop {} {\bibfield  {journal} {\bibinfo  {journal}
  {Physics Today}\ }\textbf {\bibinfo {volume} {50}},\ \bibinfo {pages} {34}
  (\bibinfo {year} {1997})}\BibitemShut {NoStop}%
\bibitem [{\citenamefont {Ulrich}\ \emph {et~al.}(2006)\citenamefont {Ulrich},
  \citenamefont {Johnson},\ and\ \citenamefont {Sutin}}]{ulrich2006}%
  \BibitemOpen
  \bibfield  {author} {\bibinfo {author} {\bibfnamefont {T.}~\bibnamefont
  {Ulrich}}, \bibinfo {author} {\bibfnamefont {P.~A.}\ \bibnamefont {Johnson}},
  \ and\ \bibinfo {author} {\bibfnamefont {A.~M.}\ \bibnamefont {Sutin}},\
  }\href@noop {} {\bibfield  {journal} {\bibinfo  {journal} {Journal of
  Acoustical Society of America}\ }\textbf {\bibinfo {volume} {119}},\ \bibinfo
  {pages} {1514} (\bibinfo {year} {2006})}\BibitemShut {NoStop}%
\bibitem [{\citenamefont {Gliozzi}\ \emph {et~al.}(2006)\citenamefont
  {Gliozzi}, \citenamefont {Griffa},\ and\ \citenamefont
  {Scalerandi}}]{gliozzi2006}%
  \BibitemOpen
  \bibfield  {author} {\bibinfo {author} {\bibfnamefont {A.~S.}\ \bibnamefont
  {Gliozzi}}, \bibinfo {author} {\bibfnamefont {M.}~\bibnamefont {Griffa}}, \
  and\ \bibinfo {author} {\bibfnamefont {M.}~\bibnamefont {Scalerandi}},\
  }\href@noop {} {\bibfield  {journal} {\bibinfo  {journal} {Journal of
  Acoustical Society of America}\ }\textbf {\bibinfo {volume} {120}},\ \bibinfo
  {pages} {2506} (\bibinfo {year} {2006})}\BibitemShut {NoStop}%
\bibitem [{\citenamefont {Ulrich}\ \emph {et~al.}(2008)\citenamefont {Ulrich},
  \citenamefont {Sutin}, \citenamefont {Guyer},\ and\ \citenamefont
  {Johnson}}]{ulrich2008time}%
  \BibitemOpen
  \bibfield  {author} {\bibinfo {author} {\bibfnamefont {T.}~\bibnamefont
  {Ulrich}}, \bibinfo {author} {\bibfnamefont {A.}~\bibnamefont {Sutin}},
  \bibinfo {author} {\bibfnamefont {R.}~\bibnamefont {Guyer}}, \ and\ \bibinfo
  {author} {\bibfnamefont {P.}~\bibnamefont {Johnson}},\ }\href@noop {}
  {\bibfield  {journal} {\bibinfo  {journal} {International Journal of
  Non-Linear Mechanics}\ }\textbf {\bibinfo {volume} {43}},\ \bibinfo {pages}
  {209} (\bibinfo {year} {2008})}\BibitemShut {NoStop}%
\bibitem [{\citenamefont {Prada}\ \emph {et~al.}(2002)\citenamefont {Prada},
  \citenamefont {Kerbrat}, \citenamefont {Cassereau},\ and\ \citenamefont
  {Fink}}]{Prada2002}%
  \BibitemOpen
  \bibfield  {author} {\bibinfo {author} {\bibfnamefont {C.}~\bibnamefont
  {Prada}}, \bibinfo {author} {\bibfnamefont {E.}~\bibnamefont {Kerbrat}},
  \bibinfo {author} {\bibfnamefont {D.}~\bibnamefont {Cassereau}}, \ and\
  \bibinfo {author} {\bibfnamefont {M.}~\bibnamefont {Fink}},\ }\href@noop {}
  {\bibfield  {journal} {\bibinfo  {journal} {Inverse Problems}\ }\textbf
  {\bibinfo {volume} {18}},\ \bibinfo {pages} {1761 } (\bibinfo {year}
  {2002})}\BibitemShut {NoStop}%
\bibitem [{\citenamefont {Goursolle}\ \emph {et~al.}(2008)\citenamefont
  {Goursolle}, \citenamefont {Santos}, \citenamefont {Matar},\ and\
  \citenamefont {Call\'e}}]{Goursolle2008170}%
  \BibitemOpen
  \bibfield  {author} {\bibinfo {author} {\bibfnamefont {T.}~\bibnamefont
  {Goursolle}}, \bibinfo {author} {\bibfnamefont {S.~D.}\ \bibnamefont
  {Santos}}, \bibinfo {author} {\bibfnamefont {O.~B.}\ \bibnamefont {Matar}}, \
  and\ \bibinfo {author} {\bibfnamefont {S.}~\bibnamefont {Call\'e}},\
  }\href@noop {} {\bibfield  {journal} {\bibinfo  {journal} {International
  Journal of Non-Linear Mechanics}\ }\textbf {\bibinfo {volume} {43}},\
  \bibinfo {pages} {170 } (\bibinfo {year} {2008})},\ \bibinfo {note} {11th
  International Workshop on Nonlinear Elasticity in Materials}\BibitemShut
  {NoStop}%
\bibitem [{\citenamefont {Ulrich}\ \emph {et~al.}(2007)\citenamefont {Ulrich},
  \citenamefont {Johnson},\ and\ \citenamefont {Guyer}}]{ulrich2007}%
  \BibitemOpen
  \bibfield  {author} {\bibinfo {author} {\bibfnamefont {T.}~\bibnamefont
  {Ulrich}}, \bibinfo {author} {\bibfnamefont {P.~A.}\ \bibnamefont {Johnson}},
  \ and\ \bibinfo {author} {\bibfnamefont {R.}~\bibnamefont {Guyer}},\
  }\href@noop {} {\bibfield  {journal} {\bibinfo  {journal} {Physical Review
  Letters}\ }\textbf {\bibinfo {volume} {98}},\ \bibinfo {pages} {104301}
  (\bibinfo {year} {2007})}\BibitemShut {NoStop}%
\bibitem [{\citenamefont {Zumpano}\ and\ \citenamefont
  {Meo}(2007)}]{Zumpano20073666}%
  \BibitemOpen
  \bibfield  {author} {\bibinfo {author} {\bibfnamefont {G.}~\bibnamefont
  {Zumpano}}\ and\ \bibinfo {author} {\bibfnamefont {M.}~\bibnamefont {Meo}},\
  }\href@noop {} {\bibfield  {journal} {\bibinfo  {journal} {International
  Journal of Solids and Structures}\ }\textbf {\bibinfo {volume} {44}},\
  \bibinfo {pages} {3666 } (\bibinfo {year} {2007})}\BibitemShut {NoStop}%
\bibitem [{\citenamefont {Miniaci}\ \emph {et~al.}(2015)\citenamefont
  {Miniaci}, \citenamefont {Marzani}, \citenamefont {Testoni},\ and\
  \citenamefont {De~Marchi}}]{Miniaci_Ultrasonics_PVC}%
  \BibitemOpen
  \bibfield  {author} {\bibinfo {author} {\bibfnamefont {M.}~\bibnamefont
  {Miniaci}}, \bibinfo {author} {\bibfnamefont {A.}~\bibnamefont {Marzani}},
  \bibinfo {author} {\bibfnamefont {N.}~\bibnamefont {Testoni}}, \ and\
  \bibinfo {author} {\bibfnamefont {L.}~\bibnamefont {De~Marchi}},\ }\href@noop
  {} {\bibfield  {journal} {\bibinfo  {journal} {Ultrasonics}\ }\textbf
  {\bibinfo {volume} {56}},\ \bibinfo {pages} {251} (\bibinfo {year}
  {2015})}\BibitemShut {NoStop}%
\bibitem [{\citenamefont {Bou~Matar}\ \emph {et~al.}(2009)\citenamefont
  {Bou~Matar}, \citenamefont {Li},\ and\ \citenamefont {Van
  Den~Abeele}}]{BouMatar2009}%
  \BibitemOpen
  \bibfield  {author} {\bibinfo {author} {\bibfnamefont {O.}~\bibnamefont
  {Bou~Matar}}, \bibinfo {author} {\bibfnamefont {Y.~F.}\ \bibnamefont {Li}}, \
  and\ \bibinfo {author} {\bibfnamefont {K.}~\bibnamefont {Van Den~Abeele}},\
  }\href@noop {} {\bibfield  {journal} {\bibinfo  {journal} {Applied Physics
  Letters}\ }\textbf {\bibinfo {volume} {95}} (\bibinfo {year}
  {2009})}\BibitemShut {NoStop}%
\bibitem [{REF()}]{REFSM}%
  \BibitemOpen
  \href@noop {} {\emph {\bibinfo {title} {See Supplemental
  Material}}}\BibitemShut {NoStop}%
\bibitem [{\citenamefont {Collet}\ \emph {et~al.}(2011)\citenamefont {Collet},
  \citenamefont {Ouisse}, \citenamefont {Ruzzene},\ and\ \citenamefont
  {Ichchou}}]{Collet_Ruzzene}%
  \BibitemOpen
  \bibfield  {author} {\bibinfo {author} {\bibfnamefont {M.}~\bibnamefont
  {Collet}}, \bibinfo {author} {\bibfnamefont {M.}~\bibnamefont {Ouisse}},
  \bibinfo {author} {\bibfnamefont {M.}~\bibnamefont {Ruzzene}}, \ and\
  \bibinfo {author} {\bibfnamefont {M.}~\bibnamefont {Ichchou}},\ }\href@noop
  {} {\bibfield  {journal} {\bibinfo  {journal} {International Journal of
  Solids and Structures}\ }\textbf {\bibinfo {volume} {48}},\ \bibinfo {pages}
  {2837 } (\bibinfo {year} {2011})}\BibitemShut {NoStop}%
\bibitem [{\citenamefont {Draeger}(1997)}]{Draeger1997}%
  \BibitemOpen
  \bibfield  {author} {\bibinfo {author} {\bibfnamefont {M.}~\bibnamefont
  {Draeger}, \bibfnamefont {Carsten;~Fink}},\ }\href@noop {} {\bibfield
  {journal} {\bibinfo  {journal} {Physical Review Letters}\ }\textbf {\bibinfo
  {volume} {79}} (\bibinfo {year} {1997})}\BibitemShut {NoStop}%
\bibitem [{\citenamefont {Sutin}\ \emph {et~al.}(2004)\citenamefont {Sutin},
  \citenamefont {TenCate},\ and\ \citenamefont {Johnson}}]{SutinJASA2004}%
  \BibitemOpen
  \bibfield  {author} {\bibinfo {author} {\bibfnamefont {A.~M.}\ \bibnamefont
  {Sutin}}, \bibinfo {author} {\bibfnamefont {J.~A.}\ \bibnamefont {TenCate}},
  \ and\ \bibinfo {author} {\bibfnamefont {P.~A.}\ \bibnamefont {Johnson}},\
  }\href@noop {} {\bibfield  {journal} {\bibinfo  {journal} {The Journal of the
  Acoustical Society of America}\ }\textbf {\bibinfo {volume} {116}},\ \bibinfo
  {pages} {2779} (\bibinfo {year} {2004})}\BibitemShut {NoStop}%
\bibitem [{\citenamefont {Solodov}\ \emph {et~al.}(2011)\citenamefont
  {Solodov}, \citenamefont {D{\"o}ring},\ and\ \citenamefont
  {Busse}}]{Solodov2011}%
  \BibitemOpen
  \bibfield  {author} {\bibinfo {author} {\bibfnamefont {I.}~\bibnamefont
  {Solodov}}, \bibinfo {author} {\bibfnamefont {D.}~\bibnamefont {D{\"o}ring}},
  \ and\ \bibinfo {author} {\bibfnamefont {G.}~\bibnamefont {Busse}},\
  }\href@noop {} {\bibfield  {journal} {\bibinfo  {journal} {J. Mech. Eng.}\
  }\textbf {\bibinfo {volume} {57}},\ \bibinfo {pages} {169} (\bibinfo {year}
  {2011})}\BibitemShut {NoStop}%
\bibitem [{\citenamefont {Pecorari}\ and\ \citenamefont
  {Poznic}(2006)}]{Pecorari2006}%
  \BibitemOpen
  \bibfield  {author} {\bibinfo {author} {\bibfnamefont {C.}~\bibnamefont
  {Pecorari}}\ and\ \bibinfo {author} {\bibfnamefont {M.}~\bibnamefont
  {Poznic}},\ }\href@noop {} {\bibfield  {journal} {\bibinfo  {journal}
  {Proceedings of The Royal Society A}\ }\textbf {\bibinfo {volume} {462}},\
  \bibinfo {pages} {769} (\bibinfo {year} {2006})}\BibitemShut {NoStop}%
\end{thebibliography}

%

\end{document}